\documentclass[a4paper,useAMS,usenatbib,usegraphicx]{mn2eadapt}

\usepackage[latin1]{inputenc}
\usepackage{amsmath}
\usepackage{amsfonts}
\usepackage{amssymb}
\usepackage[english]{babel}

\usepackage{times}
\usepackage{color}
\usepackage{rotating}
\usepackage[colorlinks=false,dvips]{hyperref}
\hypersetup{pdfborder=0 0 0}

\usepackage{afterpage}

\newcommand{\eg}{e.g.\ }
\newcommand{\ie}{i.e.\ }
\newcommand{\cf}{cf.\ }
\newcommand{\Msun}{\ensuremath{\textrm{M}_{\odot}}}

\newcommand{\Lsun}{\ensuremath{\textrm{L}_{\odot}}}
\newcommand{\kms}{km\hspace{0.25em}s$^{-1}$}

\newcommand{\Halpha}{H$\alpha$}

\newcommand{\OI}{\mbox{O\hspace{0.25em}{\sc i}}}

\newcommand{\CII}{\mbox{C\hspace{0.25em}{\sc ii}}}

\newcommand{\NaI}{\mbox{Na\hspace{0.25em}{\sc i}}}
\newcommand{\MgII}{\mbox{Mg\hspace{0.25em}{\sc ii}}}

\newcommand{\SII}{\mbox{S\hspace{0.25em}{\sc ii}}}

\newcommand{\SiII}{\mbox{Si\hspace{0.25em}{\sc ii}}}
\newcommand{\SiIII}{\mbox{Si\hspace{0.25em}{\sc iii}}}
\newcommand{\CaII}{\mbox{Ca\hspace{0.25em}{\sc ii}}}

\newcommand{\FeII}{\mbox{Fe\hspace{0.25em}{\sc ii}}}
\newcommand{\FeIII}{\mbox{Fe\hspace{0.25em}{\sc iii}}}
\newcommand{\CoII}{\mbox{Co\hspace{0.25em}{\sc ii}}}
\newcommand{\CoIII}{\mbox{Co\hspace{0.25em}{\sc iii}}}
\newcommand{\NiII}{\mbox{Ni\hspace{0.25em}{\sc ii}}}
\newcommand{\Fefs}{$^{56}$Fe}

\newcommand{\Nifs}{$^{56}$Ni}

\newcommand{\KE}{$E_{\textrm{kin}}$}
\newcommand{\Dm}{$\Delta m_{15}(B)$}

\newcommand{\myto}{\hspace{0.18em}--\hspace{0.18em}}

\definecolor{myblue}{rgb}{0.0,0.0,0.9}

\definecolor{myorange}{rgb}{0.9,0.6,0.0}

\definecolor{myredder}{rgb}{1.0,0.0,0.0}

\definecolor{mygreen}{rgb}{0.0,0.7,0.0}

\begin{document}

\title[Analysis of near-UV spectra of SN\,2010jn]{The UV\,/\,optical spectra of the Type Ia supernova SN\,2010jn:\\ a bright supernova with outer layers rich in iron-group elements}

\author[Hachinger et al.]{S. Hachinger$^{1,2,3}$, P. A. Mazzali$^{1,2}$, M. Sullivan$^{4,5}$, R. S. Ellis$^{6}$, K. Maguire$^{5}$, A. Gal-Yam$^{7}$,\newauthor D. A. Howell$^{8,9}$, P. E. Nugent$^{10}$, E. Baron$^{11,12,13,14}$, J. Cooke$^{15}$, I. Arcavi$^{7}$, D. Bersier$^{16}$,\newauthor B. Dilday$^{8,9}$, P. A. James$^{16}$, M. M. Kasliwal$^{17}$, S. R. Kulkarni$^{6}$, E. O. Ofek$^7$, R. R. Laher$^{18}$,\newauthor J. Parrent$^{8,19}$, J. Surace$^{18}$, O. Yaron$^{7}$, E. S. Walker$^{20}$\\
$^1$Istituto Nazionale di Astrofisica-OAPd, vicolo dell'Osservatorio 5, 35122 Padova, Italy\\
$^2$Max-Planck-Institut f\"ur Astrophysik, Karl-Schwarzschild-Str.\ 1, 85748 Garching, Germany\\
$^3$Institut für Theoretische Physik und Astrophysik, Universit\"at W\"urzburg, Emil-Fischer-Str. 31, 97074 W\"urzburg, Germany\\
$^4$School of Physics and Astronomy, University of Southampton, Southampton SO17 1BJ, UK\\
$^5$Department of Physics (Astrophysics), University of Oxford, Keble Road, Oxford OX1 3RH, UK\\
$^6$Cahill Center for Astrophysics, California Institute of Technology, Pasadena, CA 91125, USA\\
$^7$Benoziyo Center for Astrophysics, Weizmann Institute of Science, 76100 Rehovot, Israel\\
$^8$Las Cumbres Observatory Global Telescope Network, Goleta, CA 93117, USA\\
$^9$Department of Physics, University of California, Santa Barbara, CA 93106-9530, USA\\
$^{10}$Computational Cosmology Center, Lawrence Berkeley National Laboratory, 1 Cyclotron Rd., Berkeley, CA 94720, USA\\
$^{11}$Homer L. Dodge Department of Physics and Astronomy, University of Oklahoma, 440 W Brooks, Norman, OK 73019, USA\\
$^{12}$Hamburger Sternwarte, Gojenbergsweg 112, 21029 Hamburg, Germany\\
$^{13}$Computational Research Division, Lawrence Berkeley National Laboratory, 1 Cyclotron Rd, Berkeley, CA 94720, USA\\
$^{14}$Physics Department, University of California, Berkeley, CA 94720, USA\\
$^{15}$Centre for Astrophysics \& Supercomputing, Swinburne University of Technology, Mail H30, PO Box 218, Hawthorn, Victoria 3122, Australia\\
$^{16}$Astrophysics Research Institute, Liverpool John Moores University, Twelve Quays House, Egerton Wharf, Birkenhead CH41 1LD, UK\\
$^{17}$Observatories of the Carnegie Institution of Science, 813 Santa Barbara St, Pasadena, CA 91101, USA\\
$^{18}$Spitzer Science Center, California Institute of Technology, M/S 314-6, Pasadena, CA 91125, USA\\
$^{19}$6127 Wilder Lab, Department of Physics \& Astronomy, Dartmouth College, Hanover, NH 03755, USA\\
$^{20}$Scuola Normale Superiore di Pisa, Piazza dei Cavalieri 7, 56126 Pisa, Italy}

\date{arXiv v3, 2012 Dec 23. The definitive version will be available at the web site of MNRAS.}
\pubyear{2012}
\volume{}
\pagerange{}

\maketitle

\begin{abstract}
  Radiative transfer studies of Type Ia supernovae (SNe\,Ia) hold the promise of constraining both the density profile of the SN ejecta and its stratification by element abundance which, in turn, may discriminate between different explosion mechanisms and progenitor classes. Here we analyse the Type Ia SN\,2010jn (PTF10ygu) in detail, presenting and evaluating near-ultraviolet (near-UV) spectra from the \textit{Hubble Space Telescope} and ground-based optical spectra and light curves. SN\,2010jn was discovered by the Palomar Transient Factory (PTF) 15 days before maximum light, allowing us to secure a time series of four near-UV spectra at epochs from $-$10.5 to $+$4.8 days relative to $B$-band maximum. The photospheric near-UV spectra are excellent diagnostics of the iron-group abundances in the outer layers of the ejecta, particularly those at very early times.  Using the method of `Abundance Tomography' we derive iron-group abundances in SN\,2010jn with a precision better than in any previously studied SN\,Ia. Optimum fits to the data can be obtained if burned material is present even at high velocities, including significant mass fractions of iron-group elements.  This is consistent with the slow decline rate (or high `stretch') of the light curve of SN\,2010jn, and consistent with the results of delayed-detonation models. Early-phase UV spectra and detailed time-dependent series of further SNe\,Ia offer a promising probe of the nature of the SN\,Ia mechanism.
\end{abstract}


\section{Introduction}

Supernovae (SNe) play an important role in many areas of modern astrophysics. In particular, Type Ia SNe (SNe\,Ia) produce most of the iron-group elements in the cosmos \citep{iwa99}, and can be used as `standardizable candles' to probe the expansion history of the Universe \citep[\eg][]{rie98,per99,rie07,kes09,sul11,suz12}. Because of their importance, extensive efforts are under way to build a comprehensive observational and theoretical picture of SNe\,Ia (\eg \citealt{hil00,maz07,howell11,roepke11}). A primary aim is to understand the nature and diversity of the explosions and thus to place on a physical ground the empirical calibration procedures that are used to deduce the luminosity of SNe\,Ia from their light-curve shape \citep[\eg][]{phi93,phi99}. Another aim is to determine the progenitor systems of SNe\,Ia.

One way to improve our understanding of SN\,Ia physics is to analyse and interpret detailed observations through radiation transport models. A particularly useful technique is that of `Abundance Tomography' \citep{ste05}, where a time series of SN\,Ia spectra is modelled and information obtained about the density profile of the SN ejecta and its stratification in terms of element abundances. This makes it possible to describe the mode of explosion and possibly to discriminate among different progenitor scenarios. Many detailed spectral time series for SNe\,Ia have become available in the last decade, but most of them are restricted to optical wavelengths.  The ultraviolet (UV) spectrum of SNe\,Ia is shaped by iron-group elements, which dominate line-blocking and fluorescence effects \citep{kirshner93,pauldrach96,maz00}. Thus, series of UV spectra are invaluable diagnostics for iron-group abundances and ejecta densities in different zones of the SN.  These in turn are indicators for basic explosion properties such as burning efficiency or explosion energy \citep{iwa99}. Extensive work on UV spectra of SNe\,Ia has been carried out especially in the last few years. On the observational side, compilations of UV spectra have been built up and studied \citep[\eg][]{ellis08a,fol08,buf09,fol12sdss,maguire12a}, but also the properties of individual objects have been investigated in more detail \citep[\eg][]{fol12sn2009ig}. On the theoretical side, spectral modelling has been performed to understand the diagnostic utility of UV spectra better \citep[\eg][]{sau08,fol12sn2011iv,wal12}. Both observations and models indicate that even SNe\,Ia with similar optical spectra can be very different in the UV \citep{lentz00a, ellis08a, sau08, wal12}.

Here we present observations and models of the SN 2010jn (PTF10ygu), a `normal' SN\,Ia \citep[\cf][]{bra93}, which is however quite luminous and has high line velocities as well as high-velocity features (HVFs, \eg \citealt{mazzali05a}) in the spectra. The SN was discovered by the Palomar Transient Factory (PTF) only a few days after explosion, which allowed a detailed spectral time series to be obtained, including near-UV data from the \textit{Hubble Space Telescope} (\textit{HST}). The series of combined optical-UV spectra made it possible to analyse the outer and intermediate ejecta in unprecedented detail with the tomography technique. The abundance stratification was inferred based on density profiles of single-degenerate Chandrasekhar-mass explosion models. These models \citep[\cf \eg][]{hil00} assume that SNe\,Ia are explosions of accreting carbon-oxygen (CO) white dwarfs (WDs). We used both the fast-deflagration model W7 \citep{nom84w7,iwa99}, which has been shown to match average SNe\,Ia \citep{ste05,tanaka11}, and a more energetic delayed-detonation model (WS15DD3, \citealt{iwa99}). The latter model assumes that the combustion flame, which initially propagates subsonically (deflagration), becomes supersonic (detonation) at some point \citep{kho91a}. Based on the quality of the fits, we suggest that the latter model is a more realistic description of SN\,2010jn.

The paper starts with a report on the observations and a presentation of the observed spectra, which were obtained from \mbox{$-\textrm{13}$\,{}d} to $+\textrm{5}$\,{}d (time in the SN rest frame) with respect to maximum light in the rest-frame $B$ band (Section \ref{sec:observations}). Afterwards, we discuss our models. We lay out the objectives, methods and assumptions (Sections \ref{sec:models-objectives} -- \ref{sec:models-assumptions}). Synthetic spectra are presented and the inferred abundance profiles are discussed (Sections \ref{sec:tomography-10jn-w7}, \ref{sec:tomography-10jn-wdd3}). Finally, the results are summarised and conclusions are drawn (Section \ref{sec:conclusions}).

\section{Observations, data reduction and photometric properties}
\label{sec:observations}

SN\,2010jn (PTF10ygu) was discovered on 2010 October 12 UT by the PTF \citep[][]{law09,rau09} using the Palomar 48-in telescope (P48), through the citizen science project `Galaxy Zoo Supernovae' \citep{smith11}\footnote{Discoverers of PTF10ygu in `Galaxy Zoo Supernovae': Peter Woolliams, tracey, Graham Dungworth, lpspieler, John P Langridge, Elisabeth Baeten, Tomas Raudys, adam elbourne, Robert Gagliano.}. The SN was found at a magnitude of $r\sim \textrm{19.2}$ in the Sbc galaxy NGC2929, at a heliocentric redshift of $z=\textrm{0.02505}$ (CMB-frame redshift of 0.02602, yielding a distance modulus $\mu = \textrm{35.2}$\,mag\footnote{A Hubble constant $H_0= \textrm{72}$\,km\,s$^{-1}$\,Mpc$^{-1}$ is used throughout.}), and a spatial position of RA 09:37:30.3, Dec. $+$23:09:33 (J2000). Only a marginal detection was present in data from the previous two nights, and no detection in data taken on 2010 October 08.  Because of the relatively low redshift and apparent early discovery, a classification spectrum was triggered using the Gemini-N telescope and the Gemini Multi-Object Spectrograph \citep[GMOS;][]{hook04} on 2010 October 13 (programme ID GN-2010B-Q-13). This revealed an early SN\,Ia at about 15 days before maximum light.

Based on this classification, four epochs of near-UV observations of this SN were triggered with the \textit{HST} using the Space Telescope Imaging Spectrograph (STIS) as part of the cycle 18 programme 12298: `Towards a Physical Understanding of the Diversity of Type Ia Supernovae' (PI: Ellis). A spectral monitoring campaign with ground-based telescopes was also commenced.  Multi-colour light curves were obtained with the Liverpool Telescope \citep[LT;][]{ste04} using RATCam and with the Faulkes Telescope North (FTN).

An observing log of our spectra can be found in Table \ref{tab:speclog}. We used two STIS observing modes, both with a 0.2\arcsec\ slit: the G430L/CCD on all four epochs (giving coverage from $\sim$\,2900\,\AA\ out to $\sim$\,5700\,\AA), and the G230LB/CCD (with nominal coverage from $\sim$\,2000\,\AA\ to $\sim$\,3000\,\AA) on two epochs. Unfortunately, the more sensitive G230L/MAMA mode was not available as the Multi-Anode Microchannel Array (MAMA) was offline. Since the intrinsic UV flux of SNe\,Ia below $\sim$\,2750\,\AA\ is low, the G230LB/CCD observations only provide useful data redwards of this wavelength.

\begin{table*}
\caption{Log of the spectroscopic observations of SN\,2010jn\,/\,PTF10ygu.}
\label{tab:speclog}
\centering
\begin{tabular}{cccllr}
\hline
Calendar date of & Date of & Phase & Telescope & Instrument &
Exposure\\ 
observation (UT) & obs. (MJD) & (days)$^a\!$ & &
configuration & time
(s)\\
\hline
Oct 13 & $\!\!$55482.6 & $-$12.9 & Gemini-N & GMOS\,/\,B600\,/\,R400 & 450\\
Oct 16 & $\!\!$55485.0 & $-$10.5 & \textit{HST}    & STIS\,/\,G430L/CCD & 2175\\
Oct 16 & $\!\!$55485.6 & $-$10.0 & Gemini-N & GMOS\,/\,B600\,/\,R400 & 450\\
Oct 20 & $\!\!$55489.8 & $\phantom{\textrm 0}$$-$5.9 & \textit{HST}    &
STIS\,/\,G430L/CCD  & 2175\\
Oct 20 & $\!\!$55489.9 & $\phantom{\textrm 0}$$-$5.8 & \textit{HST}    &
STIS\,/\,G230LB/CCD & 4920\\
Oct 21 & $\!\!$55490.6 & $\phantom{\textrm 0}$$-$5.1 & Gemini-N & GMOS\,/\,B600\,/\,R400 
& 450\\
Oct 26 & $\!\!$55495.4 & $\phantom{\textrm 0}$$-$0.4 & \textit{HST}    &
STIS\,/\,G430L/CCD  &
2175\\
Oct 26 & $\!\!$55495.5 & $\phantom{\textrm 0}$$-$0.3 & \textit{HST}    &
STIS\,/\,G230LB/CCD &
7665\\
Oct 27 & $\!\!$55496.6 & $\phantom{\textrm 0}$$+$0.8 & Gemini-N & GMOS\,/\,B600\,/\,R400 
& 450\\
Oct 31 & $\!\!$55500.7 & $\phantom{\textrm 0}$$+$4.8 & \textit{HST} &
STIS\,/\,G430L/CCD   & 2175\\
Nov 01 & $\!\!$55501.2 & $\phantom{\textrm 0}$$+$5.3 & WHT & ISIS\,/\,R300B\,/\,R158R &
900\\
\hline
\end{tabular}
\parbox{10.6cm}{$^a$Days in the SN rest frame relative to maximum
light in the rest-frame $B$ band.}
\end{table*}

We also make use of five ground-based optical spectra, taken with Gemini-N with GMOS and the William Herschel Telescope (WHT) with the Intermediate dispersion Spectrograph and Imaging System (ISIS). With GMOS, we used the B600 grating with a central wavelength of 450\,nm, and the R400 grating with a central wavelength of 750\,nm. With ISIS, we used the R158R (red arm) and R300B (blue arm) gratings, together with the 5300 dichroic. Again, details can be found in Table \ref{tab:speclog}. At the earliest epoch we analyse ($-$12.9\,{}d), only an optical spectrum is available. At all other epochs, we use both \textit{HST} and ground-based spectra, taken within $<$\,29 hours of each other.

\subsection{Data reduction}
\label{sec:observations-datareduction}

\subsubsection{Spectra}

The treatment of the \textit{HST} spectra is described in \citet{maguire12a}. Briefly, the spectra were downloaded from the \textit{HST} archive using the on-the-fly reprocessing (OTFR) pipeline, giving fully calibrated and extracted 1D spectra, where the reduction and extraction are optimised for point sources. The OTFR pipeline uses the latest calibration files and data parameters to perform initial 2D image reduction such as image trimming, bias and dark-current subtraction, cosmic-ray rejection, and flat-fielding. It then performs wavelength and flux calibrations. We applied further cosmic-ray removal and hot-pixel masking by hand. The ground-based spectra were reduced using standard \textsc{iraf} procedures (see Acknowledgments), including bias subtraction, flat-fielding, wavelength calibration, flux calibration, and telluric-feature removal. The techniques used were similar to those described in \citet{ellis08a}.

For all epochs except the first one we combined the various spectra available (\ie one or two \textit{HST} spectra and one ground-based spectrum). We used \textit{HST} data below $\sim$\,$\textrm{5200}$\,\AA, and ground-based data above. In each case, we matched the spectra in the overlap region (redwards from \CaII\ H\&K) using the \textit{HST} spectrum as reference. The resulting four optical\,/\,UV spectra analysed here sample the SN in regular time intervals of about five days. They are publically available through the WISeREP archive\footnote{\url{http://www.weizmann.ac.il/astrophysics/wiserep/}} \citep{yaron12}. We have corrected all final \mbox{1D} spectra for Milky Way extinction using $R_V = \textrm{3.1}$, a colour excess of $E(B-V)_{\mathrm{MW}}=\textrm{0.03}$ from the dust maps of \citet{sch98}, and the Milky Way dust extinction law of \citet*[][]{car89}.

\begin{figure*}   
  \centering
  \includegraphics[angle=0,width=12.0cm]{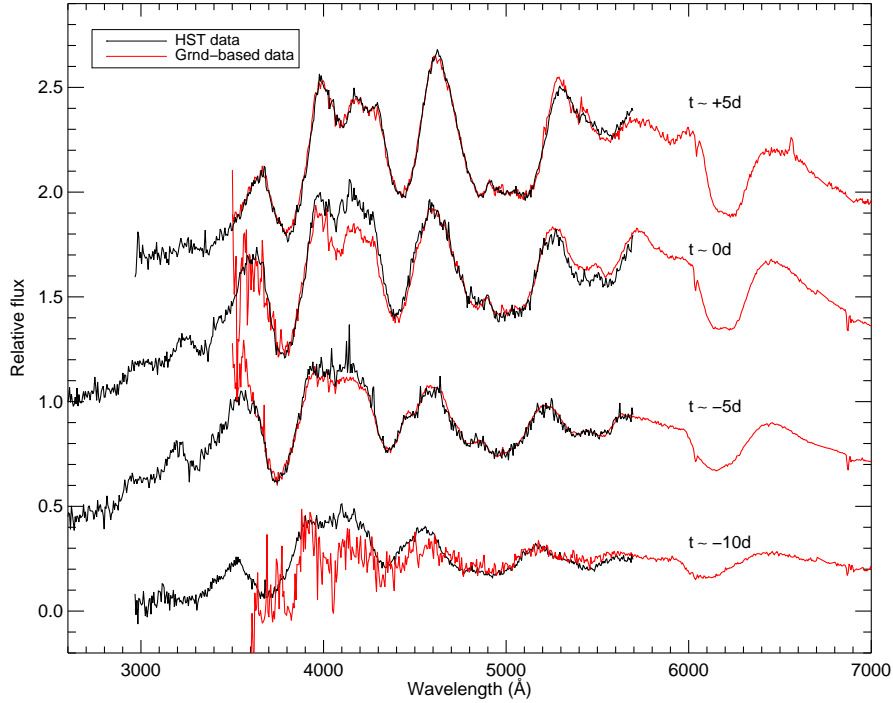}
  \caption{The importance of the \textit{HST} data (\cf Table \ref{tab:speclog}). The \textit{HST} spectra (black graphs) are compared to the Gemini-N and WHT ground-based spectra (red\,/\,grey graphs). The \textit{HST} and Gemini\,/\,WHT spectra have been scaled to each other using the overlapping regions, and arbitrary offsets have been applied on each epoch; \Halpha\ emission lines due to the host galaxy (\cf \eg Fig. \ref{fig:sequence-10jn-w7}) have been removed by inserting a constant flux for better readability. The spectra agree well in the optical, where the ground-based relative flux calibration is straightforward, and the ground-based S\,/\,N is superior to \textit{HST}. But in the blue, in general, the agreement deteriorates and the S\,/\,N of \textit{HST} becomes superior. Below 3500\,\AA, the use of \textit{HST} becomes critical.}
  \label{fig:grndspace}
\end{figure*}

We illustrate the importance of the \textit{HST} data in Fig.~\ref{fig:grndspace}, where we compare our ground-based Gemini-N\,/\,WHT and \textit{HST} spectra. Although the signal-to-noise ratio (S\,/\,N) of the ground-based spectra is generally superior in the optical at $\gtrsim$\,5000\,\AA, below 4000\,\AA\ the use of \textit{HST} becomes important as the ground-based signal usually degrades. At $\lambda<\textrm{3500}$\,\AA, the \textit{HST} data are critical. Discrepancies, sometimes significant, can be seen in the relative flux calibration of the ground and space data below 4000\,\AA. These are presumably due to the well-known difficulties with  ground-based data near the atmospheric cut-off when calibrating and correcting for atmospheric extinction.

\textit{HST} provides two other key advantages. The first is an accurate absolute flux calibration. The \textit{HST} point-spread function (PSF) is stable, allowing slit-losses to be accurately calculated and corrected for, and the overall flux calibration is believed to be accurate to a few per cent \citep{boh97}. Secondly, the host galaxy contamination in the narrow 0.2\arcsec\ slit used in the \textit{HST} observations is negligible \citep[see][]{maguire12a}; the 1\myto{}1.5\arcsec\ slits used from the ground admit more host light -- a significant potential contaminant, in particular at epochs when the SN is faint.

\subsubsection{Light curves}

Light curves were produced from the P48 $R$ and LT\,/\,FTN $gri$ photometric data using difference imaging; full details can be found in \citet{maguire12a}. In all cases, a reference image with no SN light is produced, and subtracted from each image in which the SN is present matching the PSFs of the images. In the case of the P48 data, this reference image was constructed using data from before the SN exploded; for the LT\,/\,FTN data the reference image was taken 18 months after the SN was first detected, by which time the SN flux was $<$\,1\% of its peak value.  We measure the SN photometry using a PSF fitting method. In each image frame (prior to subtraction), the PSF is determined from nearby field stars. This average PSF is then fit at the position of the SN in the difference image.  Each pixel is weighted according to Poisson statistics, yielding a SN flux and flux error.

\begin{figure*}   
  \centering
  \includegraphics[angle=0,width=12.0cm]{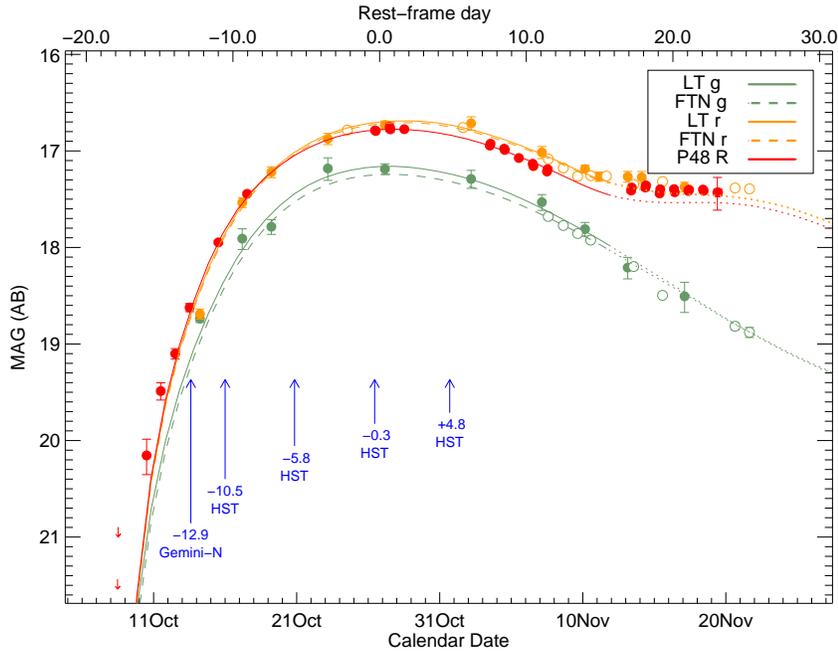}
  \caption{The optical light curves of SN\,2010jn\,/\,PTF10ygu in apparent AB magnitudes. The $R$-band photometry has been taken from the Palomar 48-in telescope (P48; solid red circles), the $g$- and $r$-band data have been obtained from the Liverpool Telescope (LT; solid green and orange circles) and the Faulkes Telescope North (FTN; open green and orange circles). No correction for extinction in the Milky Way or the host galaxy has been applied.  Over-plotted is a SiFTO light-curve fit (solid lines for P48 and LT data, and dashed lines for FTN data). Only data points up to $+$15 days past the (rest-frame) $B$-band maximum are used in the fit. The template lines are dotted after this phase and show the predicted evolution; however the SiFTO template in $R$\,/\,$r$ is not expected to reproduce the observed secondary maximum. The arrows mark the epochs (relative to the rest-frame $B$-band maximum) of the four \textit{HST} spectra.}
  \label{fig:10jn-lc}
\end{figure*}

The light curves are flux-calibrated to the Sloan Digital Sky Survey (SDSS) Data Release 8 (DR8) photometric system, close to the AB system \citep{oke83}, using stars in the field-of-view of the SN. We generally follow the procedures outlined in \citet{ofek12a}.  In the LT\,/\,FTN data ($g$, $r$ and $i$ filters), the colour terms to the SDSS system are very small. The P48 colour term is larger, and we include colour, airmass and colour--airmass terms.  The r.m.s. of the colour term fits is $\sim$\,$\textrm{0.02}$\,mag (with a colour term in $r-i$ of $\sim$\,$\textrm{0.22}$\,mag); the magnitudes presented below are, however, in the natural P48 system, \ie this colour term is not applied. Finally, we calculate aperture corrections to the SN photometry on an image-by-image basis, fitting the same PSF used for the SN flux measurement to the calibrating field stars. These corrections are small ($<$\,3\%) and account for the difference between the fixed aperture in which the calibration is performed, and the PSF fit in which the SN photometry is measured. In Fig. \ref{fig:10jn-lc} and Tables \ref{tab:10jn-lc}\,/\,\ref{tab:10jn-lc-2}, our observed light-curve data are presented.

\begin{table*}   
  \caption{Photometry of SN\,2010jn.}
  \label{tab:10jn-lc}
  \centering
  \begin{tabular}{crccccc}
    \hline
    Date of    & Phase        & $R$   & $g$   & $r$   & $i$   & Telescope \\ 
    obs. (MJD) & (days)$^a\!\!\!$ & (AB mag) & (AB mag) & (AB mag) & (AB mag) & \\
    \hline
      55477.48 &  $-$17.9 & $>$\,20.99$^b$  &    &    &    & P48  \\
      55477.52 &  $-$17.8 & $>$\,21.44$^b$  &    &    &    & P48  \\
      55479.51 &  $-$15.9 & 20.15 $\pm$ 0.18 &    &    &    & P48  \\
      55480.50 &  $-$14.9 & 19.49 $\pm$ 0.09 &    &    &    & P48  \\
      55481.51 &  $-$13.9 & 19.10 $\pm$ 0.05 &    &    &    & P48  \\
      55482.52 &  $-$13.0 & 18.62 $\pm$ 0.04 &    &    &    & P48  \\
      55483.24 &  $-$12.3 &    & 18.74 $\pm$ 0.04 &    &    & LT  \\
      55483.24 &  $-$12.3 &    &    & 18.69 $\pm$ 0.05 &    & LT  \\
      55483.24 &  $-$12.3 &    &    &    & 18.44 $\pm$ 0.02 & LT  \\
      55484.53 &  $-$11.0 & 17.95 $\pm$ 0.02 &    &    &    & P48  \\
      55486.19 &  $-$9.4  &    & 17.91 $\pm$ 0.11 &    &    & LT  \\
      55486.20 &  $-$9.4  &    &    & 17.54 $\pm$ 0.05 &    & LT  \\
      55486.20 &  $-$9.4  &    &    &    & 17.78 $\pm$ 0.07 & LT  \\
      55486.54 &  $-$9.0  & 17.44 $\pm$ 0.02 &    &    &    & P48  \\
      55488.23 &  $-$7.4  &    & 17.79 $\pm$ 0.08 &    &    & LT  \\
      55488.23 &  $-$7.4  &    &    & 17.22 $\pm$ 0.06 &    & LT  \\
      55488.23 &  $-$7.4  &    &    &    & 17.57 $\pm$ 0.11 & LT  \\
      55492.18 &  $-$3.5  &    & 17.18 $\pm$ 0.12 &    &    & LT  \\
      55492.18 &  $-$3.5  &    &    & 16.88 $\pm$ 0.06 &    & LT  \\
      55492.18 &  $-$3.5  &    &    &    & 17.36 $\pm$ 0.09 & LT  \\
      55493.55 &  $-$2.2  &    &    & 16.78 $\pm$ 0.01 &    & FTN  \\
      55493.55 &  $-$2.2  &    &    &    & 17.26 $\pm$ 0.02 & FTN  \\
      55494.18 &  $-$1.6  &    &    &    & 17.26 $\pm$ 0.05 & LT  \\
      55495.48 &  $-$0.3  & 16.79 $\pm$ 0.02 &    &    &    & P48  \\
      55495.53 &  $-$0.3  & 16.79 $\pm$ 0.01 &    &    &    & P48  \\
      55496.17 &  0.4   &    & 17.19 $\pm$ 0.06 &    &    & LT  \\
      55496.17 &  0.4   &    &    & 16.73 $\pm$ 0.04 &    & LT  \\
      55496.17 &  0.4   &    &    &    & 17.30 $\pm$ 0.05 & LT  \\
      55496.50 &  0.7   & 16.75 $\pm$ 0.04 &    &    &    & P48  \\
      55496.55 &  0.7   & 16.78 $\pm$ 0.03 &    &    &    & P48  \\
      55497.52 &  1.7   & 16.78 $\pm$ 0.01 &    &    &    & P48  \\
      55498.16 &  2.3   &    &    &    & 17.29 $\pm$ 0.03 & LT  \\
      55498.55 &  2.7   &    &    &    & 17.31 $\pm$ 0.02 & FTN  \\
      55501.63 &  5.7   &    &    & 16.76 $\pm$ 0.01 &    & FTN  \\
      55501.63 &  5.7   &    &    &    & 17.47 $\pm$ 0.02 & FTN  \\
      55502.19 &  6.2   &    & 17.29 $\pm$ 0.09 &    &    & LT  \\
      55502.19 &  6.2   &    &    & 16.71 $\pm$ 0.07 &    & LT  \\
      55502.19 &  6.2   &    &    &    & 17.39 $\pm$ 0.04 & LT  \\
      55503.50 &  7.5   & 16.94 $\pm$ 0.01 &    &    &    & P48  \\
      55503.55 &  7.6   & 16.93 $\pm$ 0.01 &    &    &    & P48  \\
      55504.50 &  8.5   & 16.98 $\pm$ 0.01 &    &    &    & P48  \\
      55504.55 &  8.5   & 16.99 $\pm$ 0.01 &    &    &    & P48  \\
      55505.53 &  9.5   & 17.07 $\pm$ 0.01 &    &    &    & P48  \\
      55506.50 &  10.4   & 17.13 $\pm$ 0.01 &    &    &    & P48  \\
      55506.54 &  10.5   & 17.16 $\pm$ 0.02 &    &    &    & P48  \\
      55507.14 &  11.1   &    & 17.53 $\pm$ 0.08 &    &    & LT  \\
      55507.14 &  11.1   &    &    & 17.01 $\pm$ 0.06 &    & LT  \\
      55507.14 &  11.1   &    &    &    & 17.75 $\pm$ 0.04 & LT  \\
      55507.50 &  11.4   & 17.21 $\pm$ 0.02 &    &    &    & P48  \\
      55507.54 &  11.5   & 17.19 $\pm$ 0.01 &    &    &    & P48  \\
      55507.59 &  11.5   &    & 17.68 $\pm$ 0.01 &    &    & FTN  \\
      55507.59 &  11.5   &    &    & 17.08 $\pm$ 0.01 &    & FTN  \\
      55507.59 &  11.5   &    &    &    & 17.84 $\pm$ 0.02 & FTN  \\
      55508.64 &  12.5   &    & 17.77 $\pm$ 0.01 &    &    & FTN  \\
      55508.64 &  12.5   &    &    & 17.18 $\pm$ 0.01 &    & FTN  \\
      55508.64 &  12.5   &    &    &    & 17.88 $\pm$ 0.02 & FTN  \\
      55509.64 &  13.5   &    & 17.86 $\pm$ 0.01 &    &    & FTN  \\
      55509.64 &  13.5   &    &    & 17.26 $\pm$ 0.01 &    & FTN  \\
      55509.64 &  13.5   &    &    &    & 17.90 $\pm$ 0.02 & FTN  \\     
  \hline
  \end{tabular}
  \parbox{11.5cm}{$^a$Days in the SN rest frame relative to maximum
                  light in the rest-frame $B$ band.\\
                  $^b$3$\sigma$ limits.\\[0.5cm]}\\[-0.5cm]
  -- continued on the next page (Table \ref{tab:10jn-lc-2}) --
\end{table*}

\begin{table*}   
  \caption{Photometry of SN\,2010jn, continued.}
  \label{tab:10jn-lc-2}
  \centering
  \begin{tabular}{crccccc}
    \hline
    Date of    & Phase        & $R$   & $g$   & $r$   & $i$   & Telescope \\ 
    obs. (MJD) & (days)$^a\!\!\!$ & (mag) & (mag) & (mag) & (mag) & \\
    \hline
      55510.15 &  14.0   &    & 17.81 $\pm$ 0.07 &    &    & LT  \\
      55510.16 &  14.0   &    &    & 17.19 $\pm$ 0.04 &    & LT  \\
      55510.16 &  14.0   &    &    &    & 17.92 $\pm$ 0.07 & LT  \\
      55510.55 &  14.4   &    & 17.92 $\pm$ 0.01 &    &    & FTN  \\
      55510.56 &  14.4   &    &    & 17.26 $\pm$ 0.01 &    & FTN  \\
      55510.56 &  14.4   &    &    &    & 17.93 $\pm$ 0.02 & FTN  \\
      55511.13 &  15.0   &    &    & 17.27 $\pm$ 0.05 &    & LT  \\
      55511.13 &  15.0   &    &    &    & 17.99 $\pm$ 0.06 & LT  \\
      55511.65 &  15.5   &    &    & 17.26 $\pm$ 0.02 &    & FTN  \\
      55513.14 &  16.9   &    & 18.21 $\pm$ 0.11 &    &    & LT  \\
      55513.14 &  16.9   &    &    & 17.27 $\pm$ 0.06 &    & LT  \\
      55513.14 &  16.9   &    &    &    & 18.27 $\pm$ 0.07 & LT  \\
      55513.38 &  17.2   & 17.41 $\pm$ 0.02 &    &    &    & P48  \\
      55513.42 &  17.2   & 17.38 $\pm$ 0.02 &    &    &    & P48  \\
      55513.55 &  17.3   &    &    & 17.35 $\pm$ 0.01 &    & FTN  \\
      55513.55 &  17.3   &    & 18.20 $\pm$ 0.02 &    &    & FTN  \\
      55513.56 &  17.3   &    &    &    & 17.85 $\pm$ 0.03 & FTN  \\
      55514.12 &  17.9   &    &    & 17.27 $\pm$ 0.06 &    & LT  \\
      55514.12 &  17.9   &    &    &    & 17.81 $\pm$ 0.02 & LT  \\
      55514.37 &  18.1   & 17.38 $\pm$ 0.03 &    &    &    & P48  \\
      55514.42 &  18.2   & 17.36 $\pm$ 0.02 &    &    &    & P48  \\
      55515.37 &  19.1   & 17.44 $\pm$ 0.02 &    &    &    & P48  \\
      55515.42 &  19.1   & 17.40 $\pm$ 0.01 &    &    &    & P48  \\
      55515.57 &  19.3   &    & 18.50 $\pm$ 0.02 &    &    & FTN  \\
      55515.57 &  19.3   &    &    & 17.32 $\pm$ 0.01 &    & FTN  \\
      55515.58 &  19.3   &    &    &    & 17.80 $\pm$ 0.02 & FTN  \\
      55516.38 &  20.1   & 17.40 $\pm$ 0.02 &    &    &    & P48  \\
      55516.42 &  20.1   & 17.43 $\pm$ 0.01 &    &    &    & P48  \\
      55517.11 &  20.8   &    & 18.51 $\pm$ 0.16 &    &    & LT  \\
      55517.12 &  20.8   &    &    & 17.37 $\pm$ 0.05 &    & LT  \\
      55517.12 &  20.8   &    &    &    & 17.83 $\pm$ 0.07 & LT  \\
      55517.38 &  21.1   & 17.41 $\pm$ 0.02 &    &    &    & P48  \\
      55517.42 &  21.1   & 17.40 $\pm$ 0.02 &    &    &    & P48  \\
      55518.40 &  22.0   & 17.40 $\pm$ 0.02 &    &    &    & P48  \\
      55518.45 &  22.1   & 17.41 $\pm$ 0.02 &    &    &    & P48  \\
      55519.41 &  23.0   & 17.43 $\pm$ 0.17 &    &    &    & P48  \\
      55519.45 &  23.1   & 17.43 $\pm$ 0.04 &    &    &    & P48  \\
      55520.16 &  23.8   &    &    &    & 17.80 $\pm$ 0.10 & LT  \\
      55520.64 &  24.2   &    & 18.82 $\pm$ 0.04 &    &    & FTN  \\
      55520.64 &  24.2   &    &    & 17.38 $\pm$ 0.07 &    & FTN  \\
      55520.65 &  24.2   &    &    &    & 17.68 $\pm$ 0.02 & FTN  \\
      55521.65 &  25.2   &    & 18.88 $\pm$ 0.05 &    &    & FTN  \\
      55521.65 &  25.2   &    &    & 17.39 $\pm$ 0.01 &    & FTN  \\
      55521.65 &  25.2   &    &    &    & 17.65 $\pm$ 0.02 & FTN  \\
  \hline
  \end{tabular}
  \parbox{11.5cm}{$^a$Days in the SN rest frame relative to maximum
                  light in the rest-frame $B$ band.}
\end{table*}

\subsection{Photometric properties}
\label{sec:observations-lcprop}

Estimating the peak luminosity of SNe\,Ia in a given rest-frame bandpass requires an interpolation between observed data points at the time of maximum light, followed by a $k$-correction \citep[\eg][]{nugent02,hsiao07} back to the standard rest-frame filter of interest using a multi-epoch spectral energy distribution (SED). We fit the observed photometry for SN\,2010jn using the SiFTO light-curve fitter \citep{conley08} developed for SN\,Ia cosmology studies.

SiFTO works in flux space, manipulating an SED template and synthesising an observer-frame light curve in a set of filters from a given spectral time series at a given redshift. These template light-curves are then fit to the data, adjusting the time-axis of each of the template light curves by a common `stretch' ($s$) factor (where the input SED time-series template is defined to have $s=1$). The normalisations in the observed filters are independent, \ie the absolute colours of the SED template used in the fit are not important and do not influence the fit. The result is a set of maximum-light observer-frame magnitudes, and a stretch factor.

Once this observer-frame SiFTO fit is complete, it can be used to estimate rest-frame magnitudes in any given set of filters, provided there is equivalent observer-frame filter coverage.  This is performed by adjusting (or `warping') the template SED to have the correct observed colours from the SiFTO fit, correcting for extinction along the line of sight in the Milky Way, de-redshifting, and integrating the resultant rest-frame SED through the required filters. This process is essentially a cross-filter $k$-correction, with the advantage that all the observed data can contribute to the overall SED shape. We only use P48 and LT data (which have a wider phase coverage than the FTN data) for this colour warping; SiFTO does not handle filters that are very close in wavelength in this process (e.g., LT $g$ and FTN $g$). However, all the FTN data are used to constrain the stretch and time of maximum light.

Our SiFTO light-curve fit is shown in Fig. \ref{fig:10jn-lc}. The fit is reasonable: the $\chi^2$ of the fit is 27 for 43 degrees of freedom.  Maximum light in the rest-frame $B$ band was on MJD $\textrm{55495.8} \pm \textrm{0.1}$, or UT 2010-10-26.8, \ie the SN was discovered some 15 days before maximum light, as expected from the analysis of the first Gemini spectrum. The stretch is $\textrm{1.06}\pm\textrm{0.02}$, corresponding to \Dm\,$=$\,0.9\,mag -- the evolution of the light curve is slower than for a typical SN\,Ia. We measure peak rest-frame absolute magnitudes at the time of $B$-band maximum light of $M_B=-\textrm{17.94} \pm \textrm{0.04}$ and $M_V=-\textrm{18.36} \pm \textrm{0.02}$, and a SiFTO colour $C$ (broadly equivalent to $B-V$ at $B$-band maximum light) of $\textrm{0.42}$\,$\pm$\,$\textrm{0.03}$. We take the $BV$ filter responses from \citet{bes90}, and these magnitudes are in the Vega system, converted from the (near)-AB system in which the photometry is measured. The uncertainties are based on statistical errors propagated through the light-curve fit. This indicates that PTF10ygu was fainter than its light-curve stretch would indicate, and redder than a normal SN\,Ia, both presumably the result of extinction by dust (see Sec.  \ref{sec:models-assumptions}).

We can estimate the ($B$-band) rise time of the SN ($\tau_r$, defined here as the time from explosion to $B$ maximum) from the very early P48 data, independently of the SiFTO light-curve fit. To this end, we measure the explosion epoch by fitting the P48 $R$ data at early epochs ($\lesssim \textrm{5}$\,{}d after discovery) using the analytical equation $f(\tau)=\alpha \times (\tau+\tau_r)^2$ \citep[`fireball model';][]{goldhaber98,rie99}. Here, $\tau=\mathfrak{t}_\textrm{o}/(\textrm{1}+z)$ where $\mathfrak{t}_\textrm{o}$ is the (observer-frame) epoch relative to $B$ maximum, and $\alpha$ is a normalising constant. This yields $\tau_{r,\textrm{fire}} = \textrm{18.6} \pm \textrm{0.3}$\,{}d. Leaving the exponent as a free fit parameter [$f(\tau)=\alpha \times (\tau+\tau_r)^\eta$], we obtain a best-fitting rise time of $\tau_{r,\textrm{free}} = \textrm{19.1} \pm \textrm{1.2}$\,{}d with a best-fitting exponent of $\eta=\textrm{2.3} \pm \textrm{0.6}$. This exponent differs from the value of $\eta=\textrm{1.8}$ recently given for `average' SNe Ia by \citet{piro12a}; such variations are, however, expected and reflect differences in the \Nifs\ distribution among the objects \citep{piro12b}. Our light-curve template (stretch 1.0) has a rise time of $\sim \textrm{17}$\,d. Scaling it to the stretch of SN\,2010jn (\ie\ to the higher opacity in such a slowly declining object), we obtain an estimate of $\tau_{r,\textrm{LCfit}} \sim \textrm{18.2}$\,d, slightly shorter than the values inferred from the early rise (and indicating the very early-time SiFTO templates may not match the data well). Our spectral models, on the other hand, lead us to prefer a somewhat higher value ($\sim \textrm{20}$\,d; Sec. \ref{sec:models-assumptions}, Appendix \ref{app:risetime}), indicating some tension between the rise time inferred from fitting the early-time data with $\eta = \textrm{2}$, and the rise time favoured by spectral modelling.

\section{Spectra and models}
\label{sec:models}

\subsection{Objectives}
\label{sec:models-objectives}

\begin{figure}   
   \centering
   \hspace*{-0.3cm}
   \includegraphics[angle=270,width=8.7cm]
   {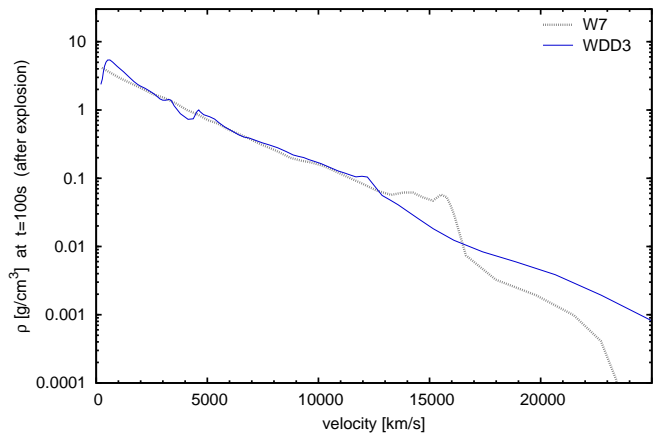}
   \caption{W7 and WS15DD3 (short WDD3) density profiles \citep{iwa99} used for
our tomography experiments.} 
   \label{fig:densitymodels}
\end{figure}

With the aim of obtaining a comprehensive picture of SN\,2010jn, we now derive ejecta models from the observed spectra. We use the techniques of Abundance Tomography \citep{ste05} to analyse the photospheric data: an ejecta model is set up, where the density distribution is that of an explosion model, but the abundances are left as `free' parameters. Synthetic spectra are then calculated from this abundance\,/\,density model and the abundances are optimised so as to match the observations with the synthetic spectra as closely as possible (\cf Sec. \ref{sec:models-methods-tomography}). Our code (Sec. \ref{sec:models-methods-mccode}) was successfully applied to obtain spectral models for single UV spectra of various SNe\,Ia \citep{sau08,wal12}. We profit from this experience here, and perform for the first time tomography of a SN\,Ia including near-UV spectra, deriving a detailed abundance stratification. The near-UV spectra, especially those taken as early as $-$10.5 or $-$5.8\,d, when the outer layers leave their imprint in the spectrum, allow us to constrain iron-group abundances in the outer and intermediate layers of the SN. We perform the tomography for different explosion models. Thus, the quality of the spectral fits can be compared. Furthermore, for each explosion model we can check the consistency of the explosive nucleosynthesis, as calculated in the hydrodynamic model, with the abundances we derive from the spectral fits. Finally, we can judge which hydrodynamic model offers a more realistic representation of SN\,2010jn.

Two different explosion models are tested in this paper: the `fast deflagration' model W7 \citep{nom84w7,iwa99} and the `delayed-detonation' model WS15DD3 \citep{iwa99}. Both represent single-degenerate Chandrasekhar-mass explosions. Their density distributions are shown in Fig.\,\ref{fig:densitymodels}.  The W7 model has been shown to provide excellent fits to light curves and spectra of average or somewhat subluminous SNe\,Ia \citep[\eg][]{ste05,mazzali08a,tanaka11}. Although its physical foundations are known not to be fully realistic, as it assumes that a deflagration flame with parametrised propagation speed burns the star, its density distribution is similar to most explosion models, except for the outermost layers, so we use it as a benchmark. WS15DD3, a delayed-detonation model, is more physically consistent, and it has a higher explosion kinetic energy and \Nifs\ mass, both of which turn out to be more appropriate for the (intrinsically) luminous SN\,2010jn.

\subsection{Methods}
\label{sec:models-methods}

We briefly describe the code which we use to calculate the synthetic spectra and the method of Abundance Tomography.

\subsubsection{Monte Carlo radiative transfer code}
\label{sec:models-methods-mccode}

We use a Monte Carlo radiative transfer code (\citealt{abb85}, \citealt{maz93}, \citealt{luc99}, \citealt{maz00} and \citealt{ste05}) in spherical symmetry to compute the formation of the spectrum in an expanding SN envelope at a given epoch. The code assumes a sharp photosphere from which radiation is emitted, and computes the interaction of the photons with the expanding SN ejecta (`atmosphere'). An initial density profile, usually the outcome of a hydrodynamic calculation, is used to calculate the densities within the atmosphere at the epoch required. This takes advantage of the fact that the ejecta are in a state of homologous, force-free expansion at the epochs relevant here (e.g. \citealt{roe05}), \ie, for each particle within the ejecta, \,$r$\,{}$=$\,{}$v$\,{}$\times$\,{}$t$\,{} is a good approximation, where $r$ is the distance from the centre, $t$ the time from explosion and $v$ the velocity of homologous expansion. Radius and velocity can therefore be used interchangeably as spatial coordinates.

At the photosphere, which is characterised by a velocity $v_{\textrm{ph}}$, the outward-flowing radiation is assumed to be a Planck continuum at a temperature $T_{\textrm{ph}}$ [$I_{\nu}^{+}$\,{} = \,{}$B_{\nu} (T_{\textrm{ph}})$]. Although this is a relatively rough approximation \citep{bongard08}, especially at late times and in SNe where line opacity dominates (\eg \citealt{hachinger09}), and it can result in a mismatch in the overall flux in the red and infrared, it has only minor consequences on the determination of the abundances from prominent spectral lines.

We simulate the propagation of the emitted photons through the envelope, considering `photon packets' which undergo Thomson scattering and bound-bound processes. The latter processes are treated in the Sobolev approximation, and a good approximation to the bound-bound emissivity is ensured by a downward branching scheme.  The construction of the simulation enforces radiative equilibrium \citep{luc99}. When packets are scattered back into the photosphere, they are considered to be re-absorbed and do not contribute to the output luminosity. An essential feature of the code is the consistent determination of a stationary state of the radiation field and of the excitation\,/\,ionisation structure of the plasma. To this end, a modified nebular approximation \citep{maz93,maz00} is used to calculate the gas state from a radiation temperature $T_\textrm{R}$ and a dilution factor $W$ cell per cell. The gas state and the radiation field are iterated in turn until the $T_\textrm{R}$ values within the atmosphere are converged to the per cent level. The temperature at the photosphere, $T_\textrm{ph}$, is automatically adjusted in these cycles so as to match a given bolometric output luminosity $L_\textrm{bol}$, compensating for the re-absorption of packets. Finally, the emerging spectrum is not calculated from packet counts, which would introduce additional `Monte Carlo noise', but rather via a formal integral, using a source function from the packet statistics \citep{luc99}.

\subsubsection{Abundance Tomography}
\label{sec:models-methods-tomography}

For an assumed density profile we model our series of early-phase spectra, inferring an optimum abundance stratification. The modelling is performed as in `Abundance Tomography' studies [where, however, also the innermost layers are studied using nebular spectra and light curves \citep[\eg][]{ste05}]. The analysis is based on the idea that the opaque core of the expanding ejecta shrinks with time (in $v$ space). In other words, the photosphere recedes, and deeper and deeper layers become visible and influence the spectral lines.

From the earliest spectrum available we can obtain a photospheric velocity and abundances within the outer ejecta such that the synthetic spectrum matches the observed one. The next observed spectrum will still show absorption by the material in the outer zone, but additional layers, which only now are above the photosphere, also contribute to the absorption. Keeping the previously inferred abundance values in the outer layers, the abundances in the newly exposed layers and a new photospheric velocity can be inferred. We continue this with later spectra. In some cases, abundances in the outer layers need to be revised in order to optimise the fit to later spectra. In this case the entire spectral sequence is consistently re-computed.

\subsection{Some general characteristics of SN\,2010jn}
\label{sec:models-10jn-characteristics}

\begin{table}
\caption{Velocities of HVFs in the combined optical\,/\,near-UV spectra of SN\,2010jn\,/\,PTF10ygu. The values given for each feature correspond to the blueshift of the point of deepest absorption. In \SII\ $\lambda$5640 this point normally corresponds to absorption near the photosphere, while in \SiII\ $\lambda$6355 and the \CaII\ IR feature it is determined by a high-velocity component before maximum. In the later spectra, a photospheric and a HVF component can be distinguished in \SiII\ $\lambda$6355.}
\label{tab:hvf}
\centering
\begin{tabular}{rccc}
\hline
Phase  & v(\CaII\ $\lambda$\,$\sim$\,$\textrm{8600}$)  & v(\SiII\ $\lambda$6355) & v(\SII\ $\lambda$5640) \\ 
(days)$^{a}\!\!\!\!$ & (\kms)          &  (\kms) & (\kms) \\
\hline
 $-$12.9 &  35000     &  23000$\phantom{^b}$ & 19000$^b$\\
 $-$10.5 &  28000     &  20000$\phantom{^b}$ & 18000$^b$\\
 $-$5.8 &   22000     &  18000$\phantom{^b}$ & 13000$^b$\\
 $-$0.3 &   19000     &  17500\,/\,14000$^b$ & 12000$^b$\\
 $+$4.8 &   17000     &  17000\,/\,13500$^b$ & 11500$^b$\\
\hline
\end{tabular}
\parbox{7.28cm}{$^a$With respect to the rest-frame $B$-band maximum; for combined spectra: epoch of the HST spectrum (\cf Tab. \ref{tab:speclog}).\\
$^b$Values refer to photospheric absorption.}
\end{table}

SN\,2010jn had a very broad light curve, corresponding to a high \Nifs\ mass, but it was a spectroscopically normal SN\,Ia \citep[\cf][]{bra93}. It is, apart from some details, a close analogue of SN\,1999ee [\citet{hamuy02b}; see maximum-light optical spectra in Fig.  \ref{fig:99ee-comparison}] rather than of SN\,1991T \citep[\eg][]{filippenko92b} or SN\,1999aa \citep{garavini04}, all of which, however, share similar light-curve properties.

The UV flux in SN\,2010jn is low, as in other SNe\,Ia, indicating significant line blocking. The P-Cygni features due to the \CaII\ IR triplet and \SiII\ $\lambda$6355 (Fig. \ref{fig:99ee-comparison}) show very strong HVFs (\eg \citealt{mazzali05a}), with shapes similar to SN\,1984A \citep{barbon89}. Although the high-velocity components are blended with the photospheric absorption in SN\,2010jn, we have been able to measure extremely high absorption velocities in the lines (Table \ref{tab:hvf}). The corresponding absorption occurs far above the photosphere (\cf the photospheric velocities of our models, Sections  \ref{sec:tomography-10jn-w7} and \ref{sec:tomography-10jn-wdd3}). Only around $B$ maximum do the HVFs vanish.

Both \SiIII\ $\lambda$4560 and \SiII\ $\lambda$5972 are very weak or absent in SN\,2010jn (Fig. \ref{fig:99ee-comparison}). The \SiIII\ $\lambda$4560 feature is characteristic for SNe\,Ia with high temperature, while \SiII\ $\lambda$5972 is strongest in low-temperature objects \citep{nug95,hac08}. The weakness of both indicates a small abundance of intermediate-mass elements (IME, $\textrm{9} \leq Z \leq \textrm{20}$), which is indicative of luminous SNe \citep{maz07}. Only the intrinsically strong \SiII\ $\lambda$6355 line remains prominent.

\subsection{The rise time and reddening of SN\,2010jn}
\label{sec:models-assumptions}

\begin{figure}   
  \centering
  \hspace*{-0.1cm}
  \includegraphics[angle=270,width=8.5cm]{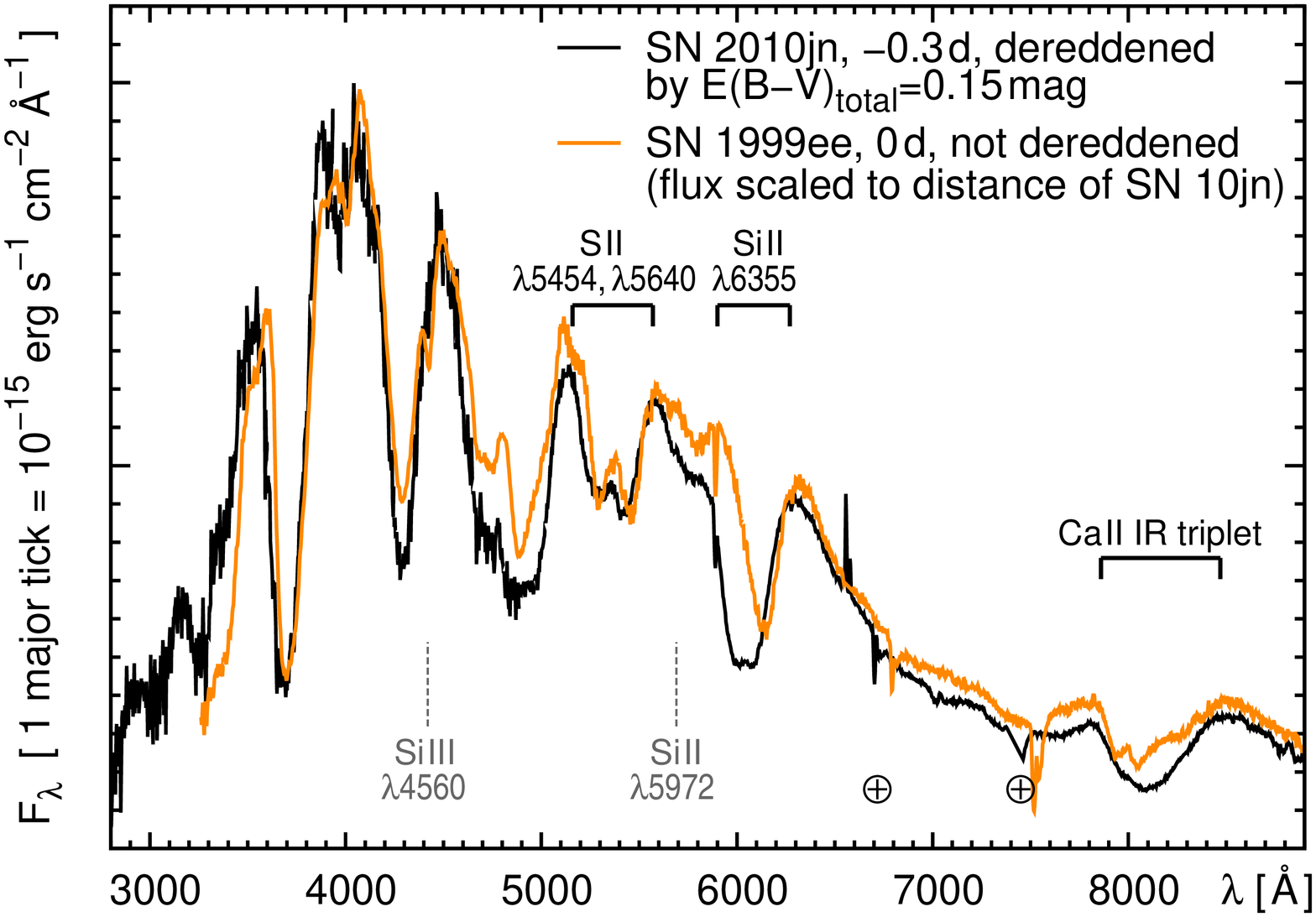}
  \caption{Spectra of SN\,2010jn (this work, black line) and SN\,1999ee \citep[][orange\,/\,grey line]{hamuy02b} around maximum light. The spectrum of SN\,1999ee is displayed as observed, but has been multiplied by a factor to account for the difference in distance towards the two SNe [\mbox{$\Delta\mu \sim \textrm{1.7}$\,mag} from NED -- \citet{mazzarella07}]. The spectrum of SN\,2010jn has been de-reddened by a total $E(B-V)$ of 0.15\,{}mag, and is then seen to coincide nicely with that of SN\,1999ee. The spectral features discussed in the text are marked (features missing in SN\,2010jn in grey with dashed marks). The positions of prominent telluric features are marked with a `$\oplus$' sign.}
  \label{fig:99ee-comparison}
\end{figure}

Since our code requires an epoch (time from explosion $t$) and a luminosity $L_\textrm{bol}$ for each model spectrum as input, we need to make assumptions about the $B$-band rise time $\tau_r$ of the SN  (\cf Sec. \ref{sec:observations-lcprop}) and the reddening.

The spectral models presented here assume a rise time of 20\,d. We have created tomography models for different values of the rise time (Appendix \ref{app:risetime}), and find that a shorter rise time would imply higher (and probably unrealistic) iron-group abundances in the outer ejecta. We conclude that after explosion the light curve of SN~2010jn has probably risen slower than predicted by a $t^2$ model. Variations in rise time and in the shape of the early light curve among SNe Ia are not entirely unexpected: all this is somewhat modulated by photon diffusion times, which depend on where \Nifs\ and other iron-group elements (which contribute to the opacity) are produced. After the submission of this work, \citet{piro12b} published a paper demonstrating such dependencies. They discuss in detail the rise of Type I SNe, supporting the view that variations in the early light-curve shape and delays in the rise are possible. All in all, our value of 20\,d for SN~2010jn is quite in line with statistical studies on the rise time of SNe\,Ia \citep{groom98,rie99,con06,strovink07}: average values of \mbox{17\myto{}20\,d} have been found for different samples, currently also with evidence for low-\Dm\ SNe (as SN\,2010jn with \Dm\,$=$\,0.9\,mag) to rise slower \citep[\eg][]{strovink07,hayden10a,ganeshalingham11}. Some objects at \Dm\,$\lesssim$\,1.0 reach values even longer than 20\,d.

The reddening to SN\,2010jn can be estimated in various ways. We have been able to measure the equivalent width of the host-galaxy \NaI\,D line in our observed low-resolution spectra (2.1\,\AA\ on average). Thus, we can estimate the host-galaxy reddening using the more conservative (lower-reddening) relation of \citet{turatto03}: \\ \[ E(B-V)\,[\textrm{mag}] = EW(\textrm{\NaI\ D})\,[\textrm{\AA}] \times \textrm{0.16} - \textrm{0.01.}\] \\ Including Galactic reddening, we arrive at a total reddening \mbox{$E(B-V)_\textrm{total} = \textrm{0.35}$\,mag}\footnote{For small redshifts, such as that of SN\,2010jn, Galactic and host extinction values can simply be added up to obtain the total reddening.} with this method. Since $EW(\textrm{\NaI\ D})$ in low-resolution spectra may be a relatively uncertain proxy for extinction \citep{blo09,poznanski11} we obtained further independent estimates.

We can infer the reddening of SN\,2010jn relative to a bright SN with a similar light-curve shape, SN\,{}1999ee [\Dm{}$=$\,{}0.95; \citet{hamuy02b}]. Comparison of the spectrum of SN\,2010jn at $-$0.2\,{d} and a maximum-light spectrum of SN\,1999ee (Fig. \ref{fig:99ee-comparison}) suggests a reddening difference of $\Delta E(B-V)_\textrm{total} \sim \textrm{0.15}$\,{}mag. \citet{sasdelli11} modelled SN\,1999ee in detail and found that it was most probably reddened by $E(B-V)_\textrm{total} = \textrm{0.26}$\,{}mag. Assuming that the SNe are almost identical, this gives a reddening of $E(B-V)_\textrm{total} \sim \textrm{0.41}$\,{}mag for SN\,2010jn. 

Finally, our light-curve fits give a peak optical colour of $B-V=\textrm{0.42}$\,{}mag (Sec.~\ref{sec:observations-lcprop}), which can also be taken as an indicator of extinction. 

As the three extinction estimates are reasonably consistent with one other, we use their average, \[E(B-V)_\textrm{total} = \textrm{0.39\,mag.} \]

\subsection{Models based on the W7 density distribution}
\label{sec:tomography-10jn-w7}

We begin by computing synthetic spectra using W7, in order to compare the results for SN\,2010jn with those obtained for other SNe [SN\,2002bo, \citet{ste05}; SN\,2003du, \citet{tanaka11}; SN\,2004eo, \citet{mazzali08a}]. Since SN\,2010jn was more luminous than all these SNe, differences in the results may be expected.

\begin{figure*}   
   \centering
   \includegraphics[width=14.5cm]{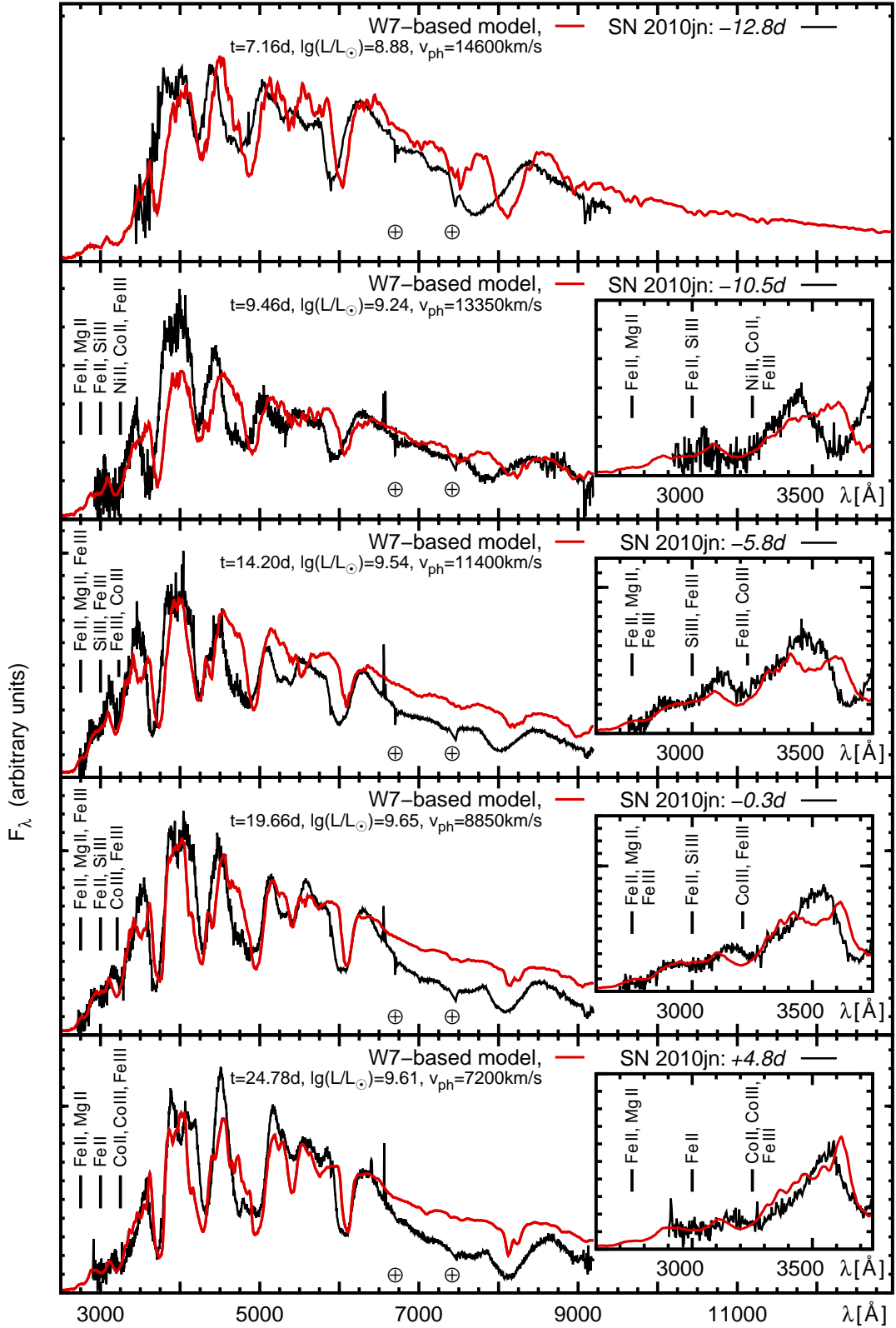}
   \caption{\mbox{10jn-W7} model sequence, based on the \mbox{W7} density profile (red\,/\,grey lines). The observed spectra are shown in black for comparison; insets show the near UV in more detail. The positions of prominent telluric features are marked with a `$\oplus$' sign. Observations and models are shown de-redshifted (\ie\ in the SN's rest frame); no correction for host-galaxy reddening has been applied to the observed spectra [the models have been reddened by $E(B-V)_\textrm{host}=\textrm{0.36}$\,{}mag]. For the most prominent UV features, the respective ions dominating the absorption in our model are given in the insets (in order of importance). Note that Ti, V and Cr cause significant quasi-continuous absorption by a large number of weak lines at $\lesssim$\,3200\,\AA, which are not reported here.}
   \label{fig:sequence-10jn-w7}
\end{figure*}

Our best-fitting synthetic spectra for SN\,2010jn based on the W7 model are shown in Fig.~\ref{fig:sequence-10jn-w7} together with the observations. We first discuss the spectral fits, and then we point out some major shortcomings of the W7-based model. Finally, we show and discuss the abundance stratification we infer using the W7 density.

\begin{figure*}   
  \centering
  \includegraphics[angle=270,width=13.5cm]{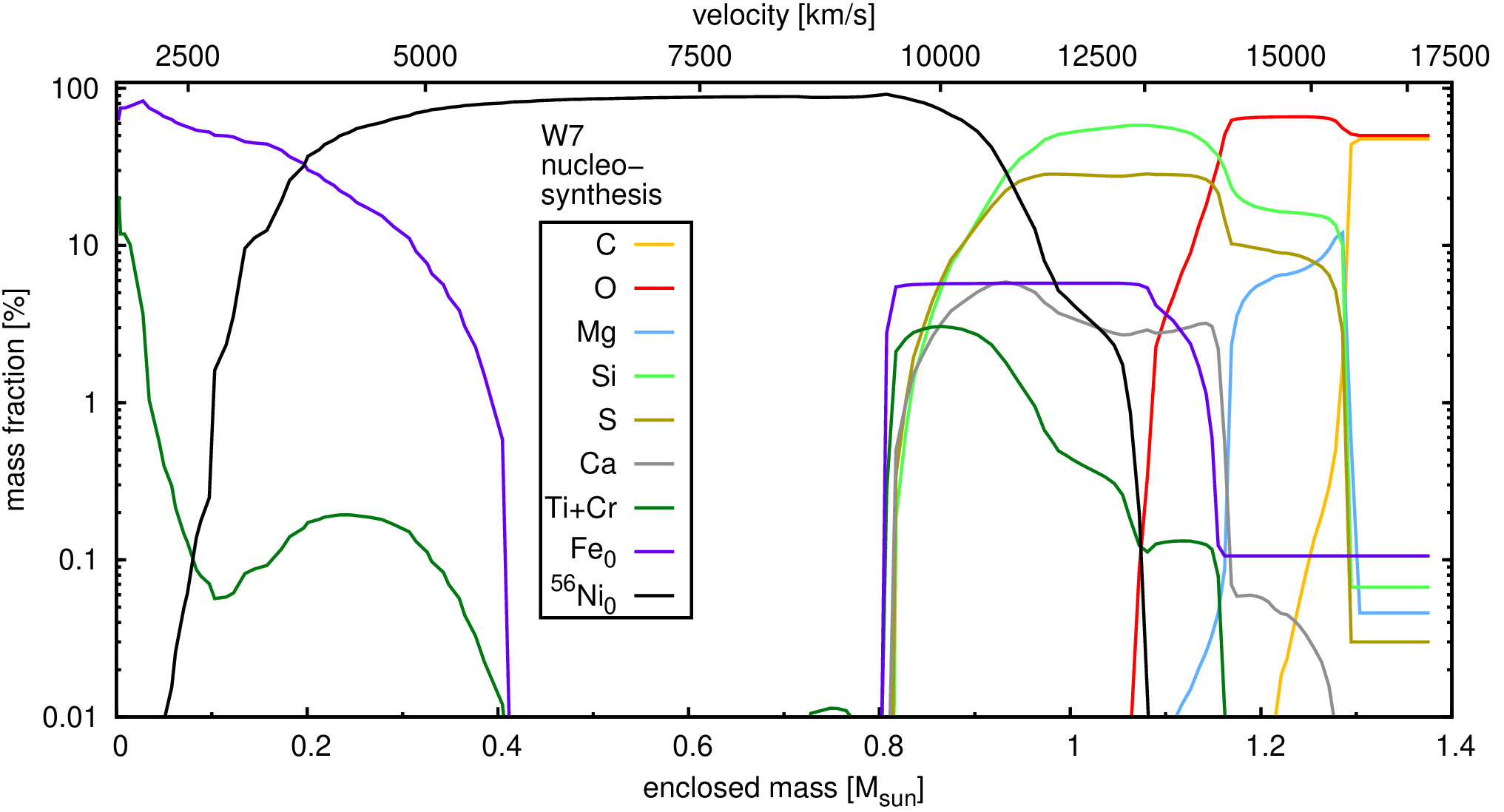}\\[0.15cm]
  \includegraphics[angle=270,width=13.5cm]{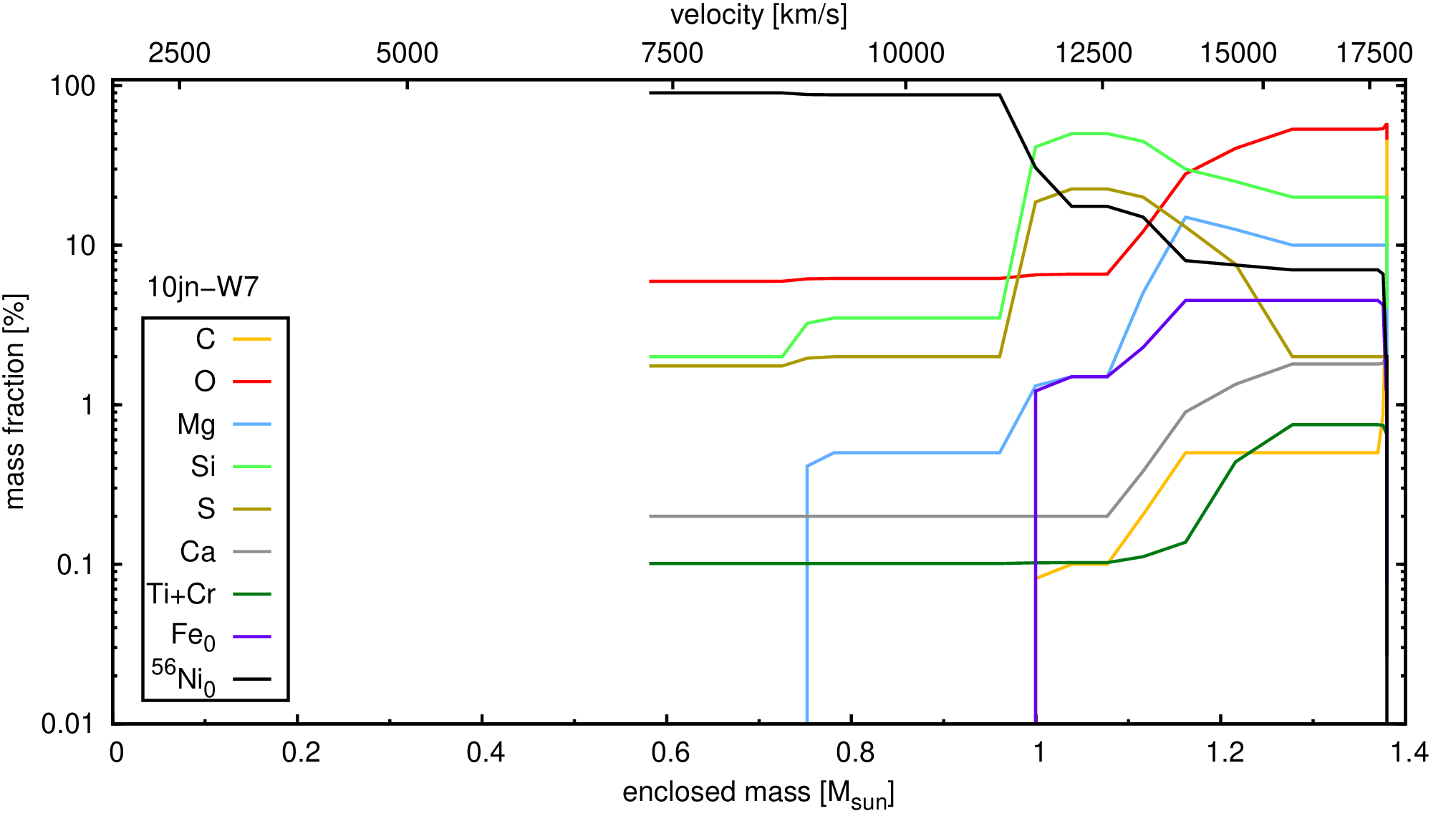}\\[-0.35cm]
  \includegraphics[angle=270,width=13.5cm]{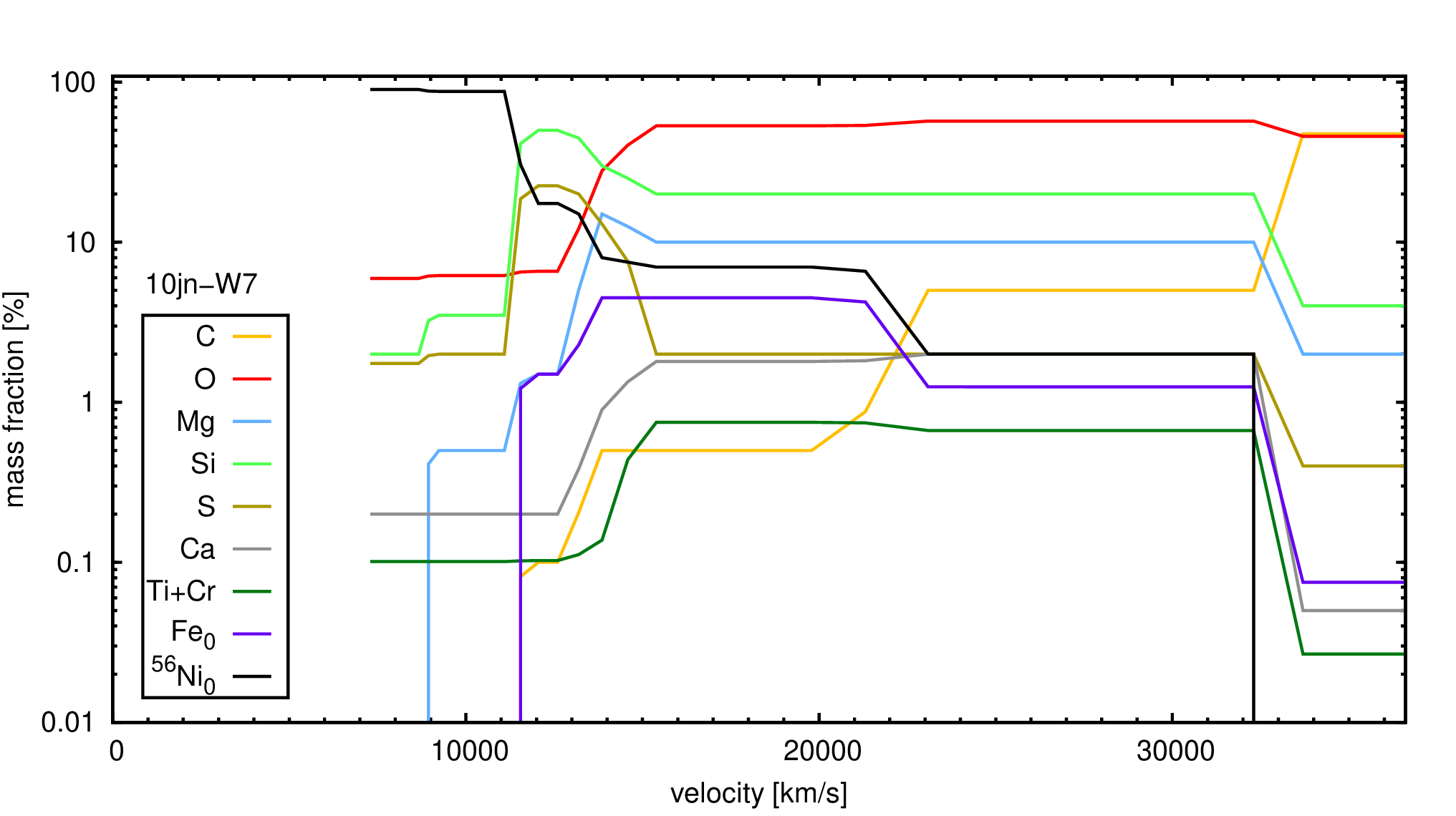}
  \caption{Abundances of W7 nucleosynthesis calculations \citep[][\textit{top
  panel}, plotted in mass space]{iwa99} vs. Abundance Tomography of SN 2010jn,
  based on the W7 density profile (\mbox{10jn-W7} model -- \textit{middle
  panel}: in mass space; \textit{lower panel}: in velocity space). The C 
  abundance in our  model is an upper limit, as no C feature is present in the
  observations. The Ni\,/\,Co\,/\,Fe abundances are given in terms of the mass
  fractions of \Nifs\ and stable Fe at $t=\textrm{0}$ [$X(^{56}\textrm{Ni}_0)$,
  $X(\textrm{Fe}_0)$]; in our spectral models, no stable Ni or Co and no
  radioactive Fe are assumed to be present.}
  \label{fig:abundances-10jn-w7}
\end{figure*}

\subsubsection{Spectral models}

The first spectrum we model has an epoch of 12.9\,d before $B$-band maximum, \ie 7.1\,d after explosion (for $t_r = \textrm{20}$\,d). The model luminosity is $L_\textrm{bol}=\textrm{0.75}\times \textrm{10}^9 \Lsun$. The spectrum shows all lines characterising a typical SN\,Ia in the photospheric phase. This indicates that not only O, but also IME ($\sim$\,34\% by mass) and some iron are present above the photosphere, which is located at 14600\,\kms\ in our model. The observed lines show large blueshifts, which cannot be reproduced by the W7-based model as it lacks high-velocity material. The absorption troughs at $\sim \textrm{4300}$\AA\ and $\sim \textrm{4700}$\AA, which are both due to a mix of Fe and IME lines, serve as diagnostics for the Fe content in the absence of UV data\footnote{The $\sim \textrm{4700}$\AA\ trough is slightly too deep in the synthetic spectrum at this epoch; however, it is slightly too shallow at the next epoch.}.

At 10.5\,{}d before $B$-band maximum, the W7-based model has a bolometric luminosity of $L_\textrm{bol}=\textrm{1.7}\times \textrm{10}^9 \Lsun$ and a photospheric velocity of \mbox{13350\,\kms}. The temperature of the photospheric black-body spectrum has risen from $T_\textrm{ph}=\textrm{11230}$\,{}K (at $-$12.9\,{}d) to $T_\textrm{ph}=\textrm{12830}$\,{}K. The low UV flux is indicative of efficient line blocking, and requires significant iron-group abundances in the photospheric layer [$X(^{56}\textrm{Ni}_0)\sim \textrm{8}$\%, $X(\textrm{Fe}_0)\sim \textrm{5}$\% and $X(\textrm{Cr})\sim \textrm{0.1}$\% at the photosphere]\footnote{The Ni\,/\,Co\,/\,Fe abundances are given in terms of the mass fractions of \Nifs\ and stable, directly synthesised Fe (mostly $^{54}$Fe) at $t=\textrm{0}$ [$X(^{56}\textrm{Ni}_0)$, $X(\textrm{Fe}_0)$]; no stable Ni or Co and no radioactive Fe are assumed to be present, as these species -- assuming realistic abundances -- have less of an impact on our models.}. Reverse-fluorescence processes, induced by iron-group elements and IME in the outer layers, actually enhance the UV flux in SNe\,Ia \citep{maz00}. In higher-luminosity models such as the ones for SN\,2010jn, however, iron-group elements also have a strong absorbing effect in the UV except for some opacity windows \citep{wal12}. Additionally, the radiation field in these models is bluer, disfavouring reverse fluorescence. The two prominent near-UV absorption features in the $-$10.5\,d model ($\sim$\,$\textrm{3000}$\,\AA\ and $\sim$\,$\textrm{3200}$\,\AA) are mostly due to \FeII\ and \NiII\,/\,\CoII\ lines, respectively. At $\lesssim$\, 2900\,\AA, the UV flux is depressed by \FeII\ and \MgII\ lines (and other lines further bluewards). As the $\sim \textrm{3200}$\,\AA\ feature is dominated by \NiII\ and \CoII\ lines, we could separate the effects of directly synthesised Fe, \Nifs\ (with decay products) and lighter iron-group elements (Ti\,/\,V\,/\,Cr) on the spectrum. As we discuss further in Sec. \ref{sec:tomography-10jn-metals}, we have thus been able to separately determine abundances of Fe and \Nifs\ and a total abundance of Ti\,/\,V\,/\,Cr in the outer ejecta.

Five days later ($- \textrm{5.8}$\,{}d, \ie $\sim$\,$\textrm{14.2}$\,d after explosion), the SN had brightened significantly ($L_\textrm{bol}=\textrm{3.5}\times \textrm{10}^9 \Lsun$), and the spectrum had become somewhat bluer (as indicated by the higher $T_\textrm{ph}=\textrm{13010}$\,{}K). The photosphere in our model recedes to 11400\,\kms, and is located in a zone which is dominated by Si, but where also iron-group material is abundant. The UV feature at $\sim$\,$\textrm{3000}$\,\AA\ now has a stronger contribution of \FeIII\ (and a large contribution by \SiIII), as the higher luminosity increases ionisation. Additionally, the feature at $\sim$\,$\textrm{3200}$\,\AA, which was dominated by \NiII\ and \CoII\ lines in the earlier spectrum, is now contaminated by \FeIII. Also, strong \CoIII\ lines begin to contribute there. Because of strong back-warming effects, Si is highly ionised and little \SiII\ remains. Therefore, the \SiII\ $\lambda$5972 absorption \citep[\cf][]{nug95,hac08} is weak. The \OI\ $\lambda$7773 feature is essentially missing for the same reason. We can also set rather stringent limits on the C content in the outer ejecta, as even small C abundances produce a visible \CII\ $\lambda$6580 line (which is not seen in the observations).

For epochs at $B$-band maximum and later, SN\,2010jn displayed a normal SN\,Ia spectrum. Our models have a bolometric luminosity $L_\textrm{bol}=\textrm{4.5}\times \textrm{10}^9 \Lsun$ at $-\textrm{0.3}$\,d and $L_\textrm{bol}=\textrm{4.0}\times \textrm{10}^9 \Lsun$ at $+\textrm{4.8}$\,d, respectively. The HVFs in \SiII\ and \CaII\ become much weaker, as it is the case in most SNe\,Ia. The model therefore fits the observations better and better with time. Even though the HVFs disappear, line velocities remain high compared to other SNe\,Ia. The photospheric velocity is \mbox{8850\,\kms} even around $B$ maximum. At $+\textrm{4.8}$\,{}d, the photosphere is at $\sim$\,$\textrm{7200}$\,\kms, and the photospheric temperature has dropped from $T_\textrm{ph}=\textrm{13330}$\,K (the value at $-\textrm{0.3}$\,d) to $T_\textrm{ph}=\textrm{12570}$\,K. High iron-group mass fractions ($\sim$\,$\textrm{90}$\%) are needed at $\textrm{7200} < v < \textrm{11400}$\,\kms\ in order to fit the low UV flux after maximum. These velocities are high for a zone dominated by iron-group material, indicating efficient \Nifs\ production and a luminous SN. The UV features after maximum are dominated by lines of \FeII\ ($\sim$\,$\textrm{3000}$\,\AA) and \CoIII\,/\,\CoII\ ($\sim$\,$\textrm{3200}$\,\AA). Flux blocking around $\sim$\,$\textrm{2900}$\,\AA\ still is mostly due to \FeII\ and \MgII\ lines.

\subsubsection{Shortcomings of the W7-based model}

The major problem of the W7-based model is that the lines in the synthetic spectra at the earlier epochs generally do not reach velocities (blueshifts) as high as seen in the observed features, although we adopted photospheric velocities as high as possible. This is a result of the steep decline of the W7 density at high velocities. The mismatch can be improved by assuming a delayed-detonation model (Sec. \ref{sec:tomography-10jn-wdd3}). We did not attempt to fit HVFs, as this would require the density and\,/\,or electron density in the outermost zone to be enhanced, probably by mixing with a circumstellar medium \citep{altavilla07a}, which is beyond the scope of this paper.

Another notable shortcoming of the models is a lack of flux around 4000\,\AA\ at $-\textrm{10.5}$\,{}d. The reasons for this mismatch at one epoch are somewhat unclear; partly, it may be due to missing high-velocity absorption in the Ca H\&K feature (which leads to a lack of re-emission around 4000\,\AA), particularly apparent at this epoch.

Later synthetic spectra tend to show an excess of flux in the IR (redwards of $\sim$\,6500\,\AA) compared to the observations. This is relatively independent of the assumed density model (\cf the WDD3-based models in Sec.  \ref{sec:tomography-10jn-wdd3}). It is rather an artefact of using a black-body spectrum at the photosphere (Sec. \ref{sec:models-methods-mccode}). The resulting mismatch is however not expected to have major effects on our abundance determination. The fact that this happens so early suggests that the black body approximation fails earlier than in most SNe\,Ia. This is probably the consequence of the presence of significant amounts of \Nifs\ near and above the photosphere even when the photospheric velocities are still high.

\subsubsection{Abundance structure}
\label{sec:tomography-10jn-w7-abundancestructure}

The spectra probe the ejecta structure from the outermost layers down to 7200\,\kms\ ($+\textrm{4.8}$\,{}d photosphere). The abundances are shown and compared to the W7 nucleosynthesis in Fig.~\ref{fig:abundances-10jn-w7}.

The model contains two zones with reduced abundances of burning products at \mbox{$\textrm{22000}\textrm{\,\kms} \leq v \leq \textrm{33000}$\,\kms} and $v>\textrm{33000}$\,\kms. These are not strictly needed to improve the fit to the spectra, but were added in order to have a structure analogous with the WDD3-based models (see Sec. \ref{sec:tomography-10jn-wdd3}). 

Moving inwards, there is a zone extending over $\sim$\,$\textrm{0.2}$\,\Msun\ ($\textrm{14000}\textrm{\,\kms} \lesssim v \lesssim \textrm{22000}$\,\kms) which is dominated by O, but already contains a significant fraction of IME (Mg, Si and S) and iron-group elements.

IME clearly dominate in the velocity range $\textrm{11500}\textrm{\,\kms} \lesssim v \lesssim \textrm{14000}$\,\kms\ (\ie over a mass range of $\lesssim$\,$\textrm{0.2}$\,\Msun). Below $\sim$\,$\textrm{11500}\,\textrm{\kms}$ (at a mass coordinate of almost $\textrm{1.0}$\,\Msun), where in most SNe\,Ia IME are still dominant \citep{maz07}, the ejecta of SN\,2010jn begin to be dominated by iron-group material. This has only been seen in very luminous SNe\,Ia \citep{maz07}.

Comparing our tomography result with the W7 nucleosynthesis calculation (Fig. \ref{fig:abundances-10jn-w7}) reveals that iron-group elements are present in SN\,2010jn at higher velocities. The higher burning efficiency is consistent with the luminosity \citep[\cf][]{stritzinger06b} and also with a high kinetic energy, as indicated by the high velocities in all spectral lines. With respect to other `normal' SNe Ia studied with the tomography technique \citep[][]{ste05,mazzali08a,tanaka11}, the SN shows a far more efficient burning. Unlike the SNe analysed by these authors, SN\,2010jn exhibits rather sharp transitions between the zones dominated by different elements (iron-group, IME, C\,/\,O). This may have to do with the fact that in an efficiently burning SN\,Ia the zone in which IME are produced and the zone where O is left are constrained to relatively narrow outer shells. Finally, SN\,2010jn, in contrast to W7, does not have an outer layer where $^{54}$Fe is produced and \Nifs\ is missing (at $M$\,$\sim$\,$\textrm{1.1}$\,\Msun\ in W7). The presence of more neutron-poor ashes (\Nifs\ in contrast to $^{54}$Fe) in SN\,2010jn would be consistent with a relatively metal-poor progenitor white dwarf \citep[\cf][]{iwa99,lentz00a}; alternatively, it may indicate multi-D effects in the abundance distribution or moderate mixing in the respective layers.

\subsection{Models based on the WDD3 density}
\label{sec:tomography-10jn-wdd3}

Having noticed the shortcomings of W7 as a backdrop density profile, we now test whether the spectra of SN\,2010jn can be better fitted using the density of a delayed-detonation model. In a delayed detonation, an initial subsonic burning phase is followed by a phase of supersonic burning which more efficiently converts the original C\,/\,O mixture into processed elements \citep{kho91a}. This type of explosion can produce a larger amount of iron-group elements as well as IME, and a higher kinetic energy, placing more material at high velocities. Although this scenario appears to be very promising for SNe\,Ia, the details of the transition from subsonic to supersonic burning are unfortunately still unclear.

Given the high luminosity of SN\,2010jn and the efficient production of Fe group elements we find in our model, we use model WS15DD3 (in short WDD3), the brightest and most energetic of the WS15DD model series of \citet{iwa99}. WDD3 synthesises 0.77\,\Msun\ of \Nifs\ and has an explosion energy \KE$ = \textrm{1.43}$\,foe.

For spectral modelling, the most important difference in the density structure of WDD3 in comparison to W7 (\cf Fig. \ref{fig:densitymodels}) is the higher density in the outer part of WDD3.

\subsubsection{Spectral models}

\begin{figure*}   
   \centering
   \includegraphics[width=14.5cm]{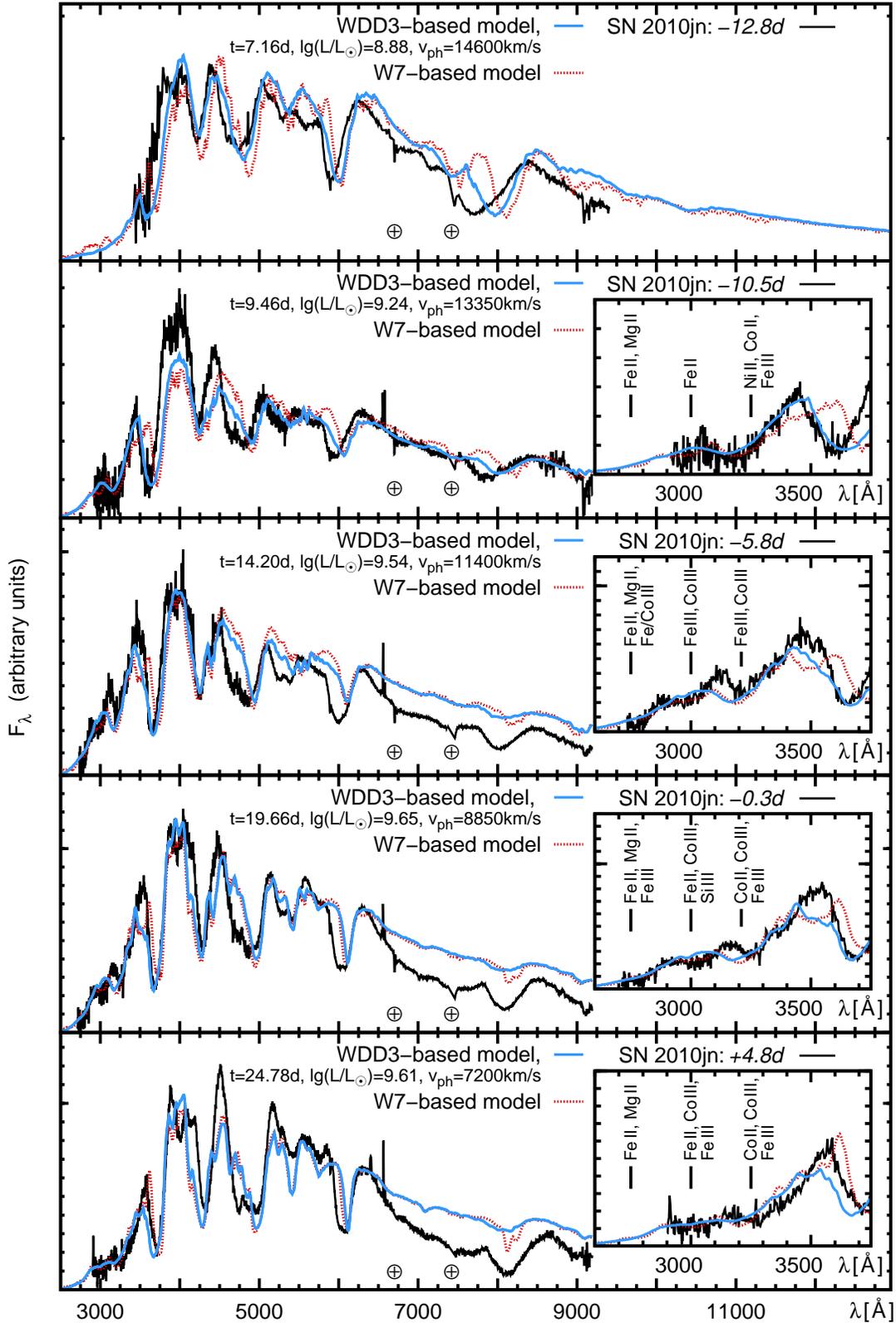}
   \caption{\mbox{10jn-WDD3} model sequence, based on the \mbox{WDD3} density profile (blue\,/\,grey, solid lines). Observed spectra (black lines) and \mbox{10jn-W7} spectra from Fig. \ref{fig:sequence-10jn-w7} (red\,/\,grey, dotted lines) are given for comparison; insets show the near UV in more detail. The differences in the density profile influence the line velocities. The positions of prominent telluric features are marked with a `$\oplus$' sign. Observations and models are shown de-redshifted (\ie\ in the SN's rest frame); no correction for host-galaxy reddening has been applied to the observed spectra [the models have been reddened by $E(B-V)_\textrm{host}=\textrm{0.36}$\,{}mag]. For the most prominent UV features, the respective ions dominating the absorption in our model are given (in order of importance). Note that Ti, V and Cr cause significant absorption by a large number of weak lines at $\lesssim$\,3200\,\AA, which are not reported here.} 
   \label{fig:sequence-10jn-wdd3}
\end{figure*}

\begin{figure*}   
  \centering
  \includegraphics[angle=270,width=13.5cm]{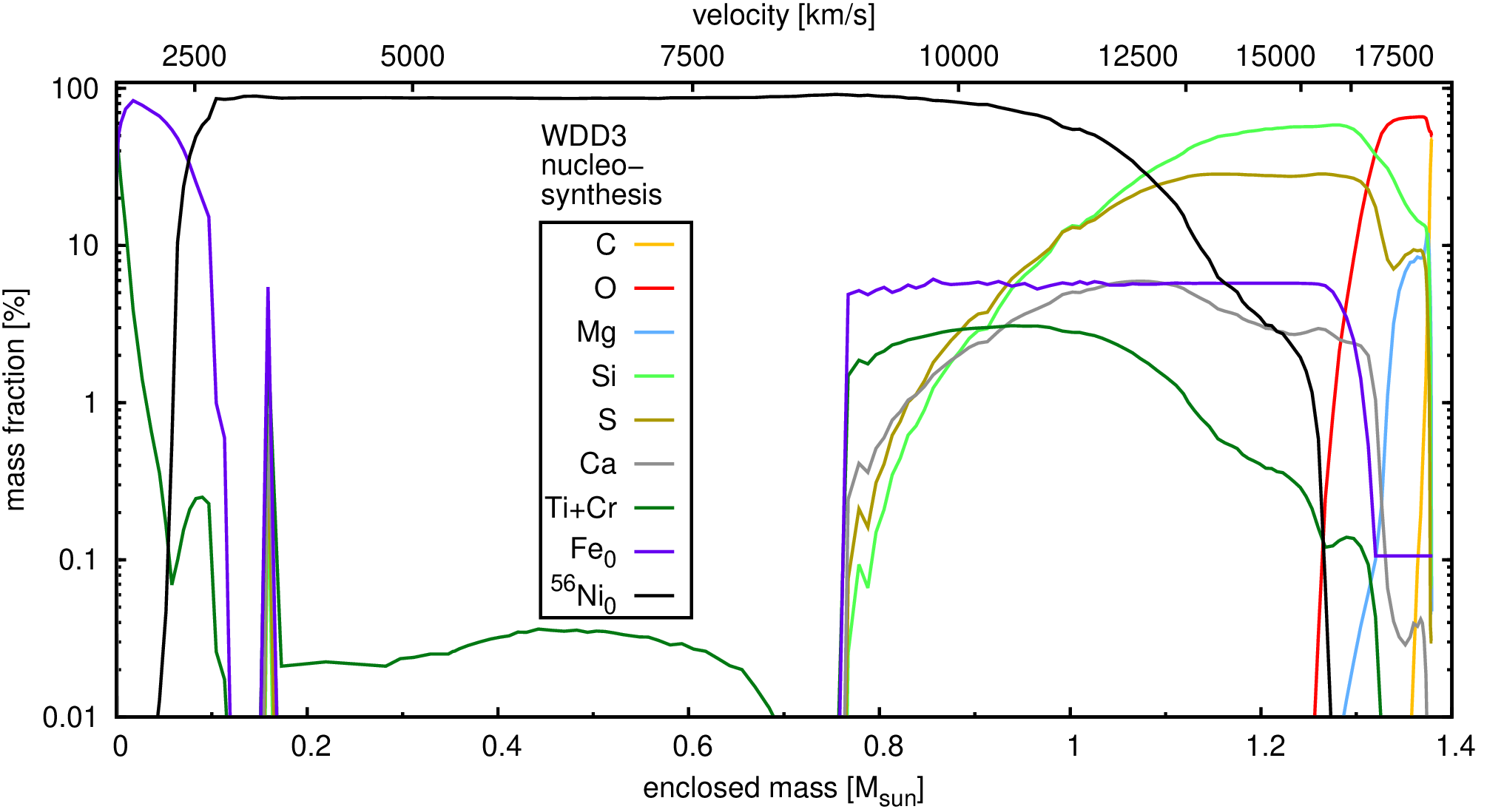}
  \\[0.15cm]
  \includegraphics[angle=270,width=13.5cm]{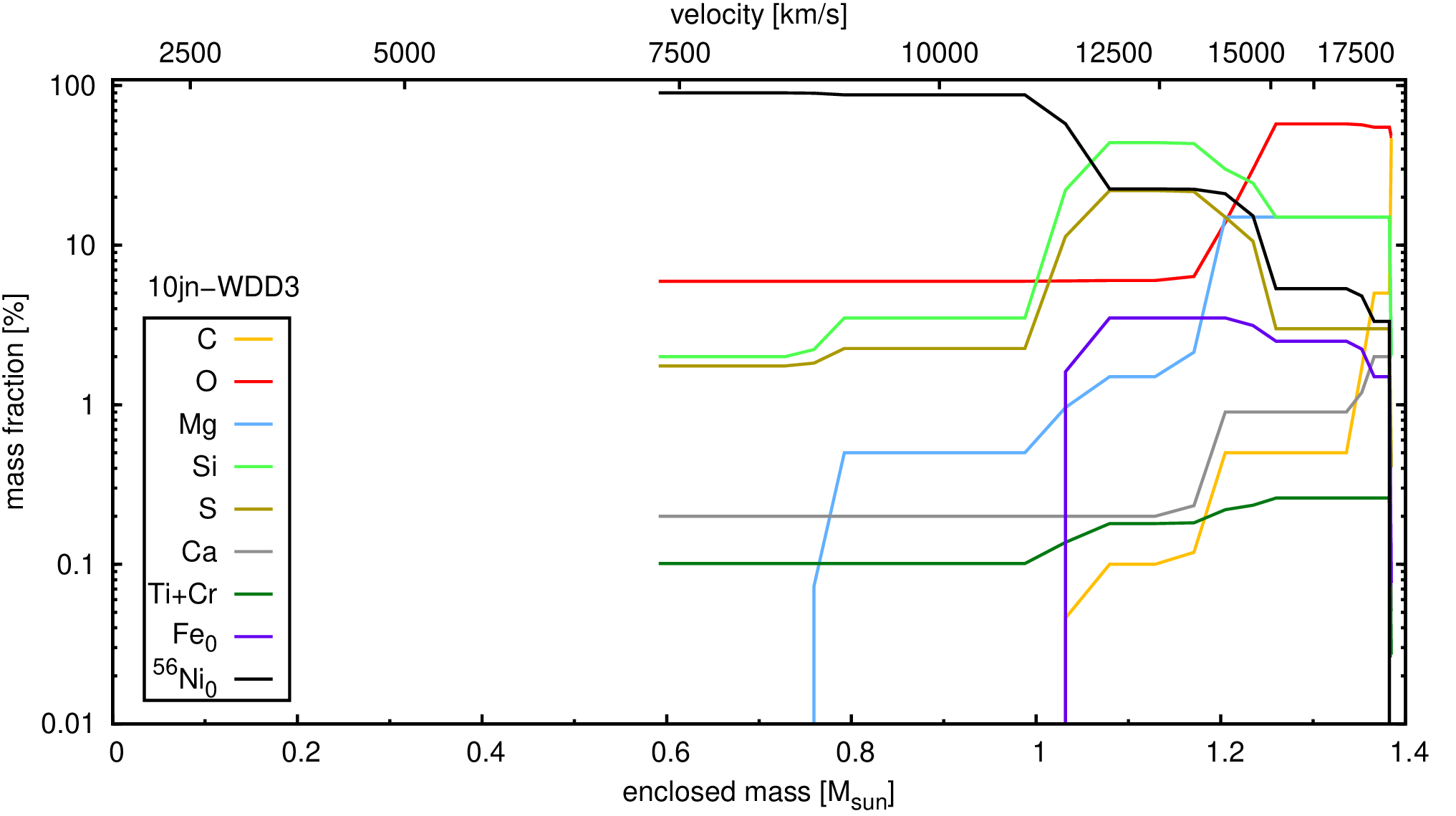}
  \\[-0.35cm]
  \includegraphics[angle=270,width=13.5cm]{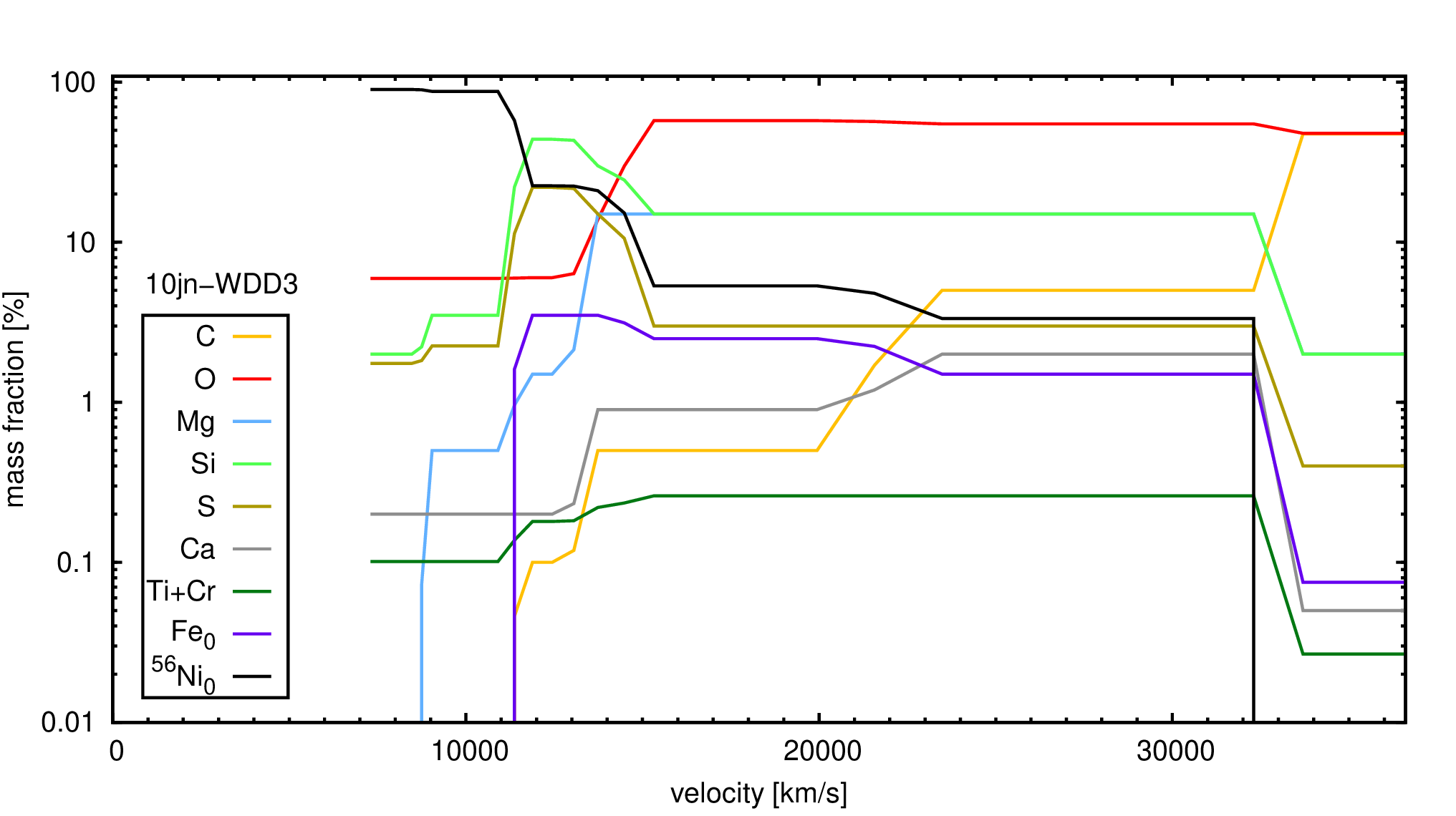}
  \caption{Abundances of WDD3 nucleosynthesis calculations \citep[][\textit{top panel}, plotted in mass space]{iwa99}, compared to our tomography based on WDD3 (\mbox{10jn-WDD3 model} -- \textit{middle panel}: in mass space; \textit{lower panel}: in velocity space). The C abundance in our model is an upper limit, as no C feature is present in the observations. The Ni\,/\,Co\,/\,Fe abundances are given in terms of the mass fractions of \Nifs\ and stable Fe at $t=\textrm{0}$ [$X(^{56}\textrm{Ni}_0)$, $X(\textrm{Fe}_0)$]; in our spectral models, no stable Ni or Co and no radioactive Fe are assumed to be present.}
  \label{fig:abundances-10jn-wdd3}
\end{figure*}

\begin{table}
\caption{Luminosities, photospheric velocities and temperatures of the
photospheric blackbody in the 10jn-WDD3 model.}
\label{tab:tvalues-10jn-wdd3}
\centering
\begin{tabular}{clcc}
\hline
 $\!\!$ Rest-frame phase $ \ \ $ & $L_\textrm{bol}$        &  $v_\textrm{ph}$ &
$T_\textrm{ph}$ \\ 
 (days)           & $\!\!\!\!$($\textrm{10}^9 \Lsun$) & (\kms)           &  (K) \\
\hline
  $-$12.9                     &   0.75  &  14600 & 10230 \\
  $-$10.5                     &   1.7   &  13350 & 11690 \\
  $\phantom{\textrm 0}$$-$5.8 &   3.5   &  11400 & 12830 \\
  $\phantom{\textrm 0}$$-$0.3 &   4.5   &  $\phantom{\textrm{0}}$8850 & 13440 \\
  $\phantom{\textrm 0}$$+$4.8 &   4.0   &  $\phantom{\textrm{0}}$7200 & 12690 \\
\hline
\end{tabular}
\end{table}

In Fig. \ref{fig:sequence-10jn-wdd3}, our WDD3-based optimum models for SN\,2010jn are compared to the data and to the W7-based models presented in the previous section.

The synthetic spectra calculated using WDD3 match the high line velocities of SN\,2010jn somewhat better [except for the HVF, which we do not attempt to reproduce -- \cf\ Sec. \ref{sec:tomography-10jn-w7} and \citet{tanaka11}]. In particular, for the earliest two epochs ($-$12.9\,d, $-$10.5\,d), \SiII\ $\lambda$6355 and the Fe-dominated trough at $\sim$\,4700\,\AA\ are broader in the WDD3-based models and give an improved fit over W7. The \CaII\ H\&K feature is significantly better matched even at later epochs; the fit quality in the near-UV region is generally similar or a bit better with WDD3. At the earliest epochs, the larger extent of the line-forming zone in WDD3 (towards high velocities) makes the WDD3-based spectra smoother and in this way more similar to the observations.

The differences with respect to W7 are basically due to the higher density in the outer zones in the delayed-detonation model, which lead to an increased absorption, also because the higher densities favour some strongly absorbing, singly ionised ion species (over doubly ionised ones). However, apart from the outer layers, the changes in the model atmospheres from W7 to WDD3 are limited. This reflects in the fact that the optimum WDD3 model could be constructed using the same values for photospheric velocity and bolometric luminosity as for W7 (cf. Table \ref{tab:tvalues-10jn-wdd3}). The photospheric black-body temperatures are similar to the W7-based models except for the first two epochs, where the WDD3-based models have a flatter temperature gradient because WDD3 has a lower density than W7 at \mbox{$\textrm{12500}\textrm{\,\kms} \leq v \leq \textrm{16000}$\,\kms} (\cf Fig. \ref{fig:densitymodels}).

The UV features in the WDD3-based model at the earliest epochs are dominated by the same elements as in the W7-based model (Fe near $\sim \textrm{3000}$\,\AA, Ni and Co near $\sim \textrm{3200}$\,\AA. At later times, all iron-group elements contribute somewhat more to all features because of increased line opacities.

\subsubsection{Abundance structure}
\label{sec:tomography-10jn-wdd3-abundancestructure}

In Fig. \ref{fig:abundances-10jn-wdd3} we show the abundances of our 10jn-WDD3 model together with a plot of the original WDD3 nucleosynthesis. This model may be regarded as a reference for SN\,2010jn: the good fits to the observed data indicate that the density of WDD3 provides a reasonable description to the actual ejecta. The abundance stratification of the WDD3-based models in velocity space is qualitatively similar to that of the W7-based models (\ref{sec:tomography-10jn-w7-abundancestructure}), despite the differences in the explosion models. Therefore, we can say that our results on the chemical structure should be reasonably independent of the density model.

In order to construct our optimum models, we have introduced two independent abundance zones in the outermost part of the ejecta (\mbox{$\textrm{22000}\textrm{\,\kms} \leq v \leq \textrm{33000}$\,\kms} and \mbox{$v\geq \textrm{33000}$\,\kms}). These zones contain fewer burning products, ensuring optimum spectral fits (\eg by avoiding spurious high-velocity absorption in the Fe\,/\,Mg feature at $\sim$\,{}4300\,\AA). The Ca, Si and S abundances at \mbox{$\textrm{22000}\textrm{\,\kms} \leq v \leq \textrm{33000}$\,\kms} have been optimised to fit the higher-velocity parts of the respective features (except for the HVFs). Despite the decrease of most burning products (with respect to the lower layers), Ca and Ti\,/\,Cr remain quite abundant, possibly as a result of incomplete burning.

Our model yields an upper limit of $\sim$\,60\% to the combined abundance of C and O at $\textrm{14600\,\kms} < v < \textrm{33000}$\,\kms. With higher mass fractions, we would have lower fractions of burning products, and this would degrade the fit.

The moderate C\,/\,O abundance between 14600 and 33000\,\kms\ is compensated by an IME mass fraction of $>$\,30\%, a mass fraction of directly synthesised Fe of $\sim$2\%, and a mass fraction of $\sim$\,5\% of \Nifs\ and decay products. The near-UV spectra make it possible to disentangle the effects of Fe, \Nifs\ and Ti\,/\,V\,/\,Cr (which we again treat combined -- \cf Sections \ref{sec:tomography-10jn-w7},\ref{sec:tomography-10jn-metals}). As in the W7-based tomography, we find \Nifs\ to be present in the outermost layers (see Fig. \ref{fig:abundances-10jn-wdd3}), and even dominating in general over directly synthesised Fe.

Assuming the WDD3 density profile, we have surveyed 0.8\Msun\ of the ejecta (down to the $+\textrm{4.8}$\,d photosphere at \mbox{7200\,\kms}). The C\,/\,O content in this region is $\sim$\,$\textrm{0.13}$\,\Msun. The mass in IME is $\sim$\,0.21\,\Msun, and $\sim$\,0.47\,\Msun\ are in iron-group elements ($\sim$\,0.45\,\Msun\ in \Nifs).

If the $\sim \textrm{0.6}$\,\Msun\ core, which cannot be analysed with our spectra, has a \Nifs\ mass fraction as at our innermost photosphere (90\%), SN\,2010jn contains almost 1\,\Msun\ of \Nifs. This is consistent with a very luminous SN\,Ia. In order to complete our analysis at the lowest velocities, nebular spectra would be required (\cf \citealt{mazzali11}).

Finally, comparing our abundance diagram with the delayed-detonation nucleosynthesis (Fig. \ref{fig:abundances-10jn-wdd3}), we find good agreement in the general structure with few exceptions. SN\,2010jn shows somewhat more mixing than WDD3: in the IME-dominated zone, both oxygen and (to a smaller degree) iron-group elements are abundant.

A main finding of this analysis is the presence of burned material in the outer layers of SN\,2010jn. The high iron-group abundances (Fe, Ti\,/\,V\,/\,Cr, \Nifs) in these zones, which lead to a high total opacity of the ejecta, explain the large light-curve stretch of SN\,2010jn. Evidence for the presence of \Nifs\ in the outer layers of SN\,Ia explosions has already been given for the less energetic SN\,2011fe \citep{nugent11,piro12a}, using different diagnostics.

\subsection{Metal abundances in the outer layers: the potential of UV spectra as diagnostics}
\label{sec:tomography-10jn-metals}

In the previous sections we have mentioned that the near-UV spectra of SN\,2010jn allowed us to determine in detail the iron-group abundances in the partially burned layers. Iron-group ratios in incompletely burned zones of SNe Ia may be characteristic for the explosion physics [\eg\ the outermost layers of sub-Chandrasekhar, edge-lit explosions may be relatively rich in Cr -- \citet{sim12}] or the metallicity of the progenitor white dwarf \citep{iwa99}.

\citet{lentz00a} and \citet{fol12sn2011iv} show that the outermost, unburned layers, where all iron-group material is from the progenitor, can have a significant effect on UV spectra. This should provide a handle for determining the pre-explosion metallicity by modelling. In SN\,2010jn, however, only the very outer layers above $\sim$\,33000\,\kms\ lack freshly synthesised iron-group material. Therefore, the near-UV\,/\,optical spectra show practically no sensitivity to iron-group elements from the progenitor.

Here, we focus on demonstrating the sensitivity of the early-time spectra on the abundances of Fe, \Nifs\ and Ti\,/\,V\,/\,Cr in the incompletely burned layers above the $-\textrm{10.5}$\,d photosphere (13350\,\kms). We do this re-computing the first two synthetic spectra (in the time series) from modified versions of our WDD3-based ejecta model, where the respective abundances are scaled up and down (Sec. \ref{sec:tomography-10jn-metals-TiFeNisequences}). Then, we calculate the entire spectral series for a model where all iron-group species are augmented\,/\,reduced simultaneously (Sec. \ref{sec:tomography-10jn-metals-FeGsequence}). Thus, we show how the overall iron-group abundance reflects in the near-UV spectra, and demonstrate that the effects due to the outer ejecta layers get weaker with time. This means that early-time UV observations of Type~Ia supernovae, which sample the incompletely burned and unburned layers, are of greatest value.

\subsubsection{Abundances of Fe, Ti\,/\,V\,/\,Cr and \Nifs\ in the outer layers: effect on the early-time spectra}
\label{sec:tomography-10jn-metals-TiFeNisequences}

\begin{figure*}   
   \centering
   \includegraphics[width=14.5cm]{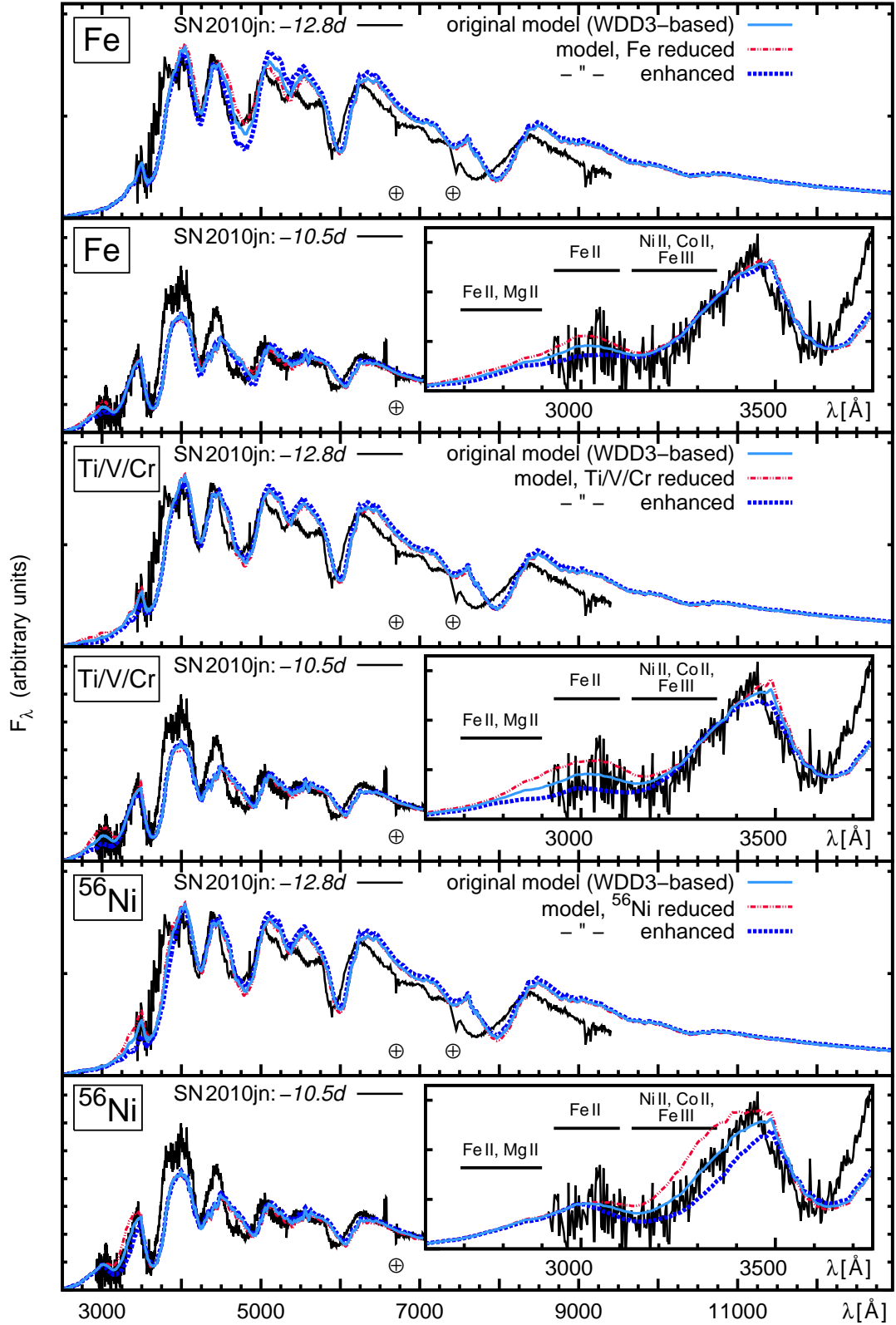}
   \caption{Sensitivity of the \mbox{10jn-WDD3} model spectra (\cf Fig. \ref{fig:sequence-10jn-wdd3}) on different iron-group elements at the earliest epochs. The abundances of the elements indicated in the different panels (at $v \gtrsim \textrm{13350}$\kms) are reduced and enhanced by a factor of two, respectively. For the most prominent UV features, the respective ions dominating the absorption are indicated in the insets; Ti\,/\,V\,/\,Cr lines are not mentioned, as the individual lines of these ions are usually weaker -- still they cause significant absorption due to their large number.}
   \label{fig:elemexperiments-10jn}
\end{figure*}

Figure \ref{fig:elemexperiments-10jn} demonstrates how the spectra allow us to determine abundances of directly synthesised Fe, Ti\,/\,V\,/\,Cr\footnote{For Ti\,/\,V\,/\,Cr we only infer a common total abundance, roughly assuming a reasonable mix of these elements of $\sim$\,1:1:10 (\eg \citealt{iwa99}).} and \Nifs\ (with decay products, depending on the epoch). It shows synthetic spectra at $-\textrm{12.9}$\,d and $-\textrm{10.5}$\,d that we have yielded by changing each abundance value in turn up and down by a factor of two. The Fe abundance mainly affects the optical feature at $\sim$\,4700\,\AA\ and the UV flux around 2900\myto{}3100\,\AA\ (however only at the second epoch, where the photosphere is well in the zone containing significant amounts of Fe). Ti, V and Cr have a similar, but stronger effect at $\sim$\,2800\myto{}3200\,\AA, and no significant influence on the optical. \Nifs, finally, affects the flux in the near-UV region at $\sim$\,3200\myto{}3400\,\AA. The behaviour of the UV features somewhat resembles that found by \citet{sau08}, who scaled the iron-group abundances up and down in one-epoch spectral models for SNe~Ia\footnote{Note that \citet{hoe98} found a flux enhancement in the UV for higher metallicity, which we do not see -- however, their modelling approach, involving a calculation of hydrodynamics\,/\,nucleosynthesis models for different metallicity, is not easily comparable to ours.}.

From optical spectra alone, the Ti\,/\,V\,/\,Cr and \Nifs\ abundances in the outer layers cannot be reliably determined: these elements have no direct influence in the optical except for fluorescence, which is difficult to predict (only at later epochs, \Fefs\ from \Nifs\ also contributes to Fe lines). UV spectra thus provide valuable additional information. Clearly, it is important in this context that the observed fluxes are reliable in the wavelength range of the relevant features. The decent flux calibration possible with \textit{HST} (\cf Sec. \ref{sec:observations-datareduction}) is thus a key advantage in the context of our study.

\subsubsection{Iron-group abundances in the outer layers: sensitivity of the spectra at different epochs}
\label{sec:tomography-10jn-metals-FeGsequence}

\begin{figure*}   
   \centering
   \includegraphics[width=14.5cm]{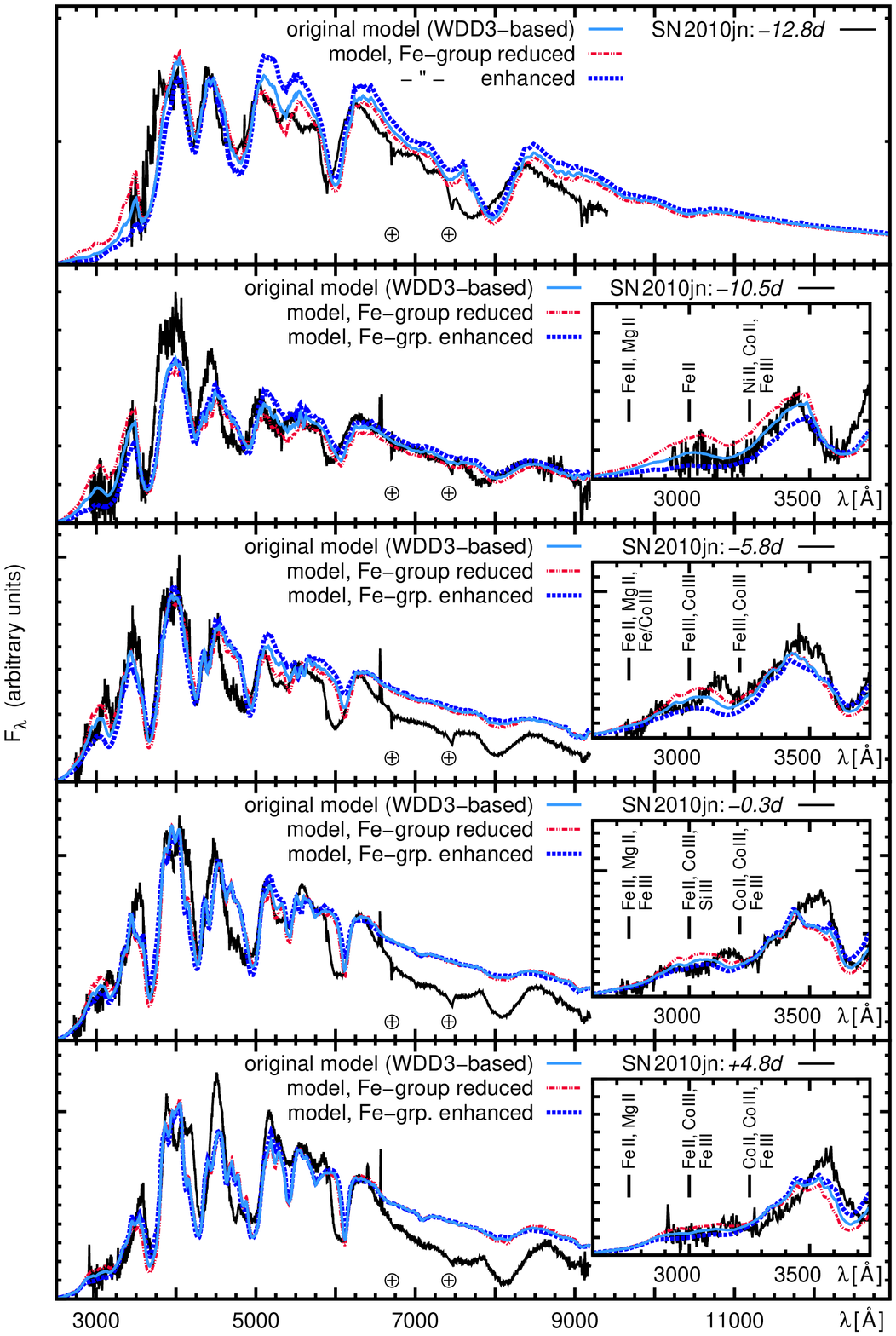}
   \caption{\mbox{10jn-WDD3} model sequence (analogous to Fig. \ref{fig:sequence-10jn-wdd3}), and two versions of this sequence where the iron-group content at $v \gtrsim \textrm{13350}$\kms\ is reduced and enhanced by a factor of two, respectively.}
   \label{fig:sequence-10jn-feexperiment}
\end{figure*}

Figure \ref{fig:sequence-10jn-feexperiment} shows the entire series of model spectra resulting when all iron-group abundances in the outer layers of the ejecta are changed by a factor of two with respect to our WDD3-based optimum models for SN\,2010jn.

The UV flux in the earliest spectra ($-\textrm{12.9}$\,d, $-\textrm{10.5}$\,d) is clearly sensitive to the iron-group content. While this flux changes by $\gtrsim \textrm{50}$\%, the variation in the optical, due to Fe lines and fluorescence effects, is smaller ($\lesssim \textrm{30}$\%).

At the later epochs (from $-\textrm{5.8}$\,d on), the UV still reacts, but somewhat less, since the photosphere is located deeper and deeper inside the region where the abundances are varied. In fact, the UV spectra around maximum light are rather sensitive to the iron-group content in the inner ejecta, which however can be better analysed using nebular-phase spectra.

\section{Summary and Conclusions}
\label{sec:conclusions}

We have for the first time used a time-series of photospheric near-UV\,/\,optical spectra of a SN\,Ia (SN\,2010jn\,/\,PTF10ygu) to analyse its abundance stratification with the tomography technique of \citet{ste05}. Our work can be seen in line with classical work aimed at understanding UV spectra of SNe Ia \citep[\eg][]{kirshner93,maz00}, with significantly refined methods for radiative transfer and spectral modelling. The early near-UV observations have proved extremely valuable for analysing the iron-group content in the outer layers. They allowed us to differentiate among different iron-group elements present in the partially burned and unburned zone. The later spectra give constraints on the abundances further inwards.

From our abundance analysis, we find that SN\,2010jn synthesised significant amounts of iron-group elements, typical for an energetic, luminous SN Ia. SN\,2010jn has only a thin oxygen-dominated zone and a limited IME zone. Even these zones contain directly synthesised Fe ($\sim \textrm{0.01}$\,\Msun) as well as \Nifs\ and decay products ($\sim \textrm{0.05}$\,\Msun\ in total). The presence of iron-group elements and intermediate-mass elements up to high velocities (even $>$\,20000\,\kms) enhances the opacity of the SN, which is consistent with the overall slow light-curve evolution.

We favour for SN\,2010jn a Chandrasekhar-mass delayed-detonation model with efficient nucleosynthesis and an explosion energy somewhat above average (WDD3, \citealt{iwa99}). It allows us to reproduce the high expansion velocities in the observed spectra better than the `fast deflagration' model W7 \citep{nom84w7,iwa99}, which has little material at high velocities. The presence of Fe and \Nifs\ in the outermost layers is not predicted by the original (1D) WDD3 model. The abundance of Fe at high velocities (\mbox{$\textrm{13350}\textrm{\,\kms} \leq v \leq \textrm{33000}$\,\kms}) is between one and two orders of magnitude above solar. This cannot come from the progenitor; it is rather a clue about the explosion properties of SN\,2010jn. A high iron-group abundance is consistent with an energetic SN, but still, explaining iron-group material in the outer layers is a challenge for explosion models. Multi-dimensional models may show some clumpiness or asymmetry which could appear to an observer as `outwards-mixing' of iron-group material.

Our models favour a longer rest-frame rise time for SN\,2010jn ($\sim$$\textrm{20}$\,d) than predicted by the empirical $t^2$ analysis of the early light curve (18.6\,d). This may imply that the SN after explosion has a `dark phase' longer than predicted by the $t^2$ (`fireball') model. It remains to be seen whether this interesting `quirk' is found in other objects that are not as luminous as SN\,2010jn. Very early data are best suited to tackle this issue.

Finally, we remark that a precise analysis of the abundances in the outer layers, as presented here, should be complemented by an analysis of the ejecta core, for which nebular spectra are necessary. Also, observations of more SNe\,Ia in the UV will be needed to establish the properties of the outermost layers of SNe\,Ia with different characteristics and light-curve properties.

\section*{ACKNOWLEDGMENTS}

This work has been made possible by the participation of more than 10,000 volunteers in the `Galaxy Zoo Supernovae' project, \href{http://supernova.galaxyzoo.org/authors}{http://supernova.galaxyzoo.org/authors}. It is based on observations made with the NASA\,/\,ESA \textit{HST}, obtained at the Space Telescope Science Institute, which is operated by the Association of Universities for Research in Astronomy, Inc., under NASA contract NAS 5-26555. The observations are associated with programme 12298. P.A.M.\ and S.H.\ acknowledge support by the programme ASI-INAF I/009/10/0. M.S.\ acknowledges support from the Royal Society, and A.G. acknowledges support by the ISF, a Minerva ARCHES award, and the Lord Sieff of Brimpton Fund. M.M.K.\ acknowledges generous support from the Hubble Fellowship and the Carnegie-Princeton Fellowship. We have used observations from the LT, operated on the island of La Palma by Liverpool John Moores University in the Spanish Observatorio del Roque de los Muchachos of the Instituto de Astrofísica de Canarias with financial support from the UK Science and Technology Facilities Council. Spectroscopic observations in the optical have been taken at the Gemini Observatory under programme ID GN-2010B-Q-13, which is operated by the Association of Universities for Research in Astronomy, Inc., under a cooperative agreement with the NSF on behalf of the Gemini partnership: the National Science Foundation (United States), the Science and Technology Facilities Council (United Kingdom), the National Research Council (Canada), CONICYT (Chile), the Australian Research Council (Australia), Minist\'{e}rio da Ci\^{e}ncia, Tecnologia e Inova\c{c}\~{a}o (Brazil) and Ministerio de Ciencia, Tecnolog\'{i}a e Innovaci\'{o}n Productiva (Argentina).  Finally, an optical spectrum has been taken at the WHT, operated on the island of La Palma by the Isaac Newton Group in the Spanish Observatorio del Roque de los Muchachos of the Instituto de Astrofísica de Canarias. We have used data from the NASA\,/\,IPAC Extragalactic Database (NED, \href{http://nedwww.ipac.caltech.edu}{http://nedwww.ipac.caltech.edu}, operated by the Jet Propulsion Laboratory, California Institute of Technology, under contract with the National Aeronautics and Space Administration). For data handling, we have made use of various software (as mentioned in the text) including \textsc{iraf}. \textsc{iraf} -- Image Reduction and Analysis Facility (\href{http://iraf.noao.edu}{http://iraf.noao.edu}) -- is an astronomical data reduction software distributed by the National Optical Astronomy Observatory (NOAO, operated by AURA, Inc., under contract with the National Science Foundation).

\bibliographystyle{mn2e}
\bibliography{diss.bib}

\appendix

\section{Choice of the $B$-band rise time}
\label{app:risetime}

\begin{figure}   
  \centering
  \includegraphics[angle=270,width=8.0cm]{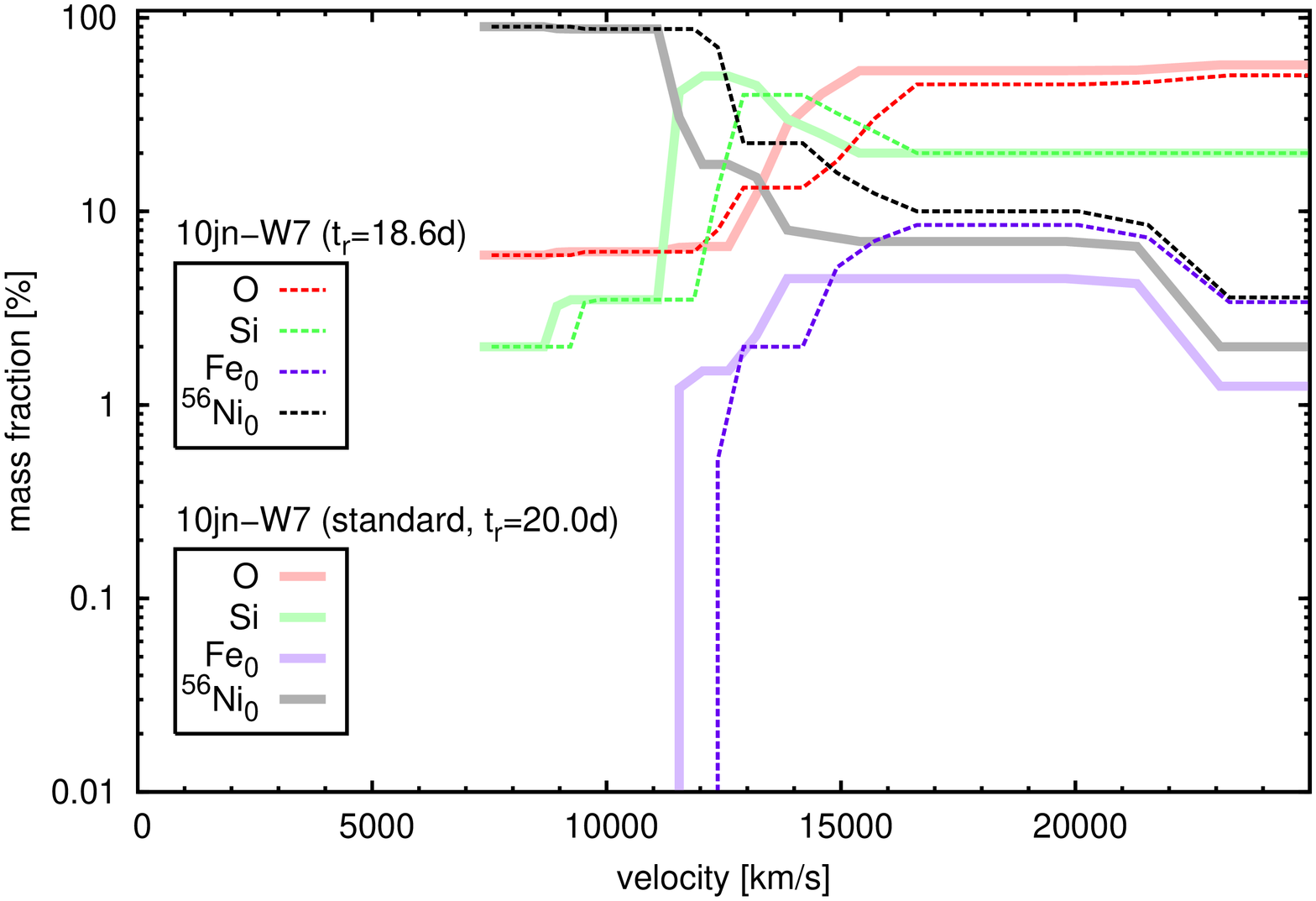}
  \\[0.35cm]
  \includegraphics[angle=270,width=8.0cm]{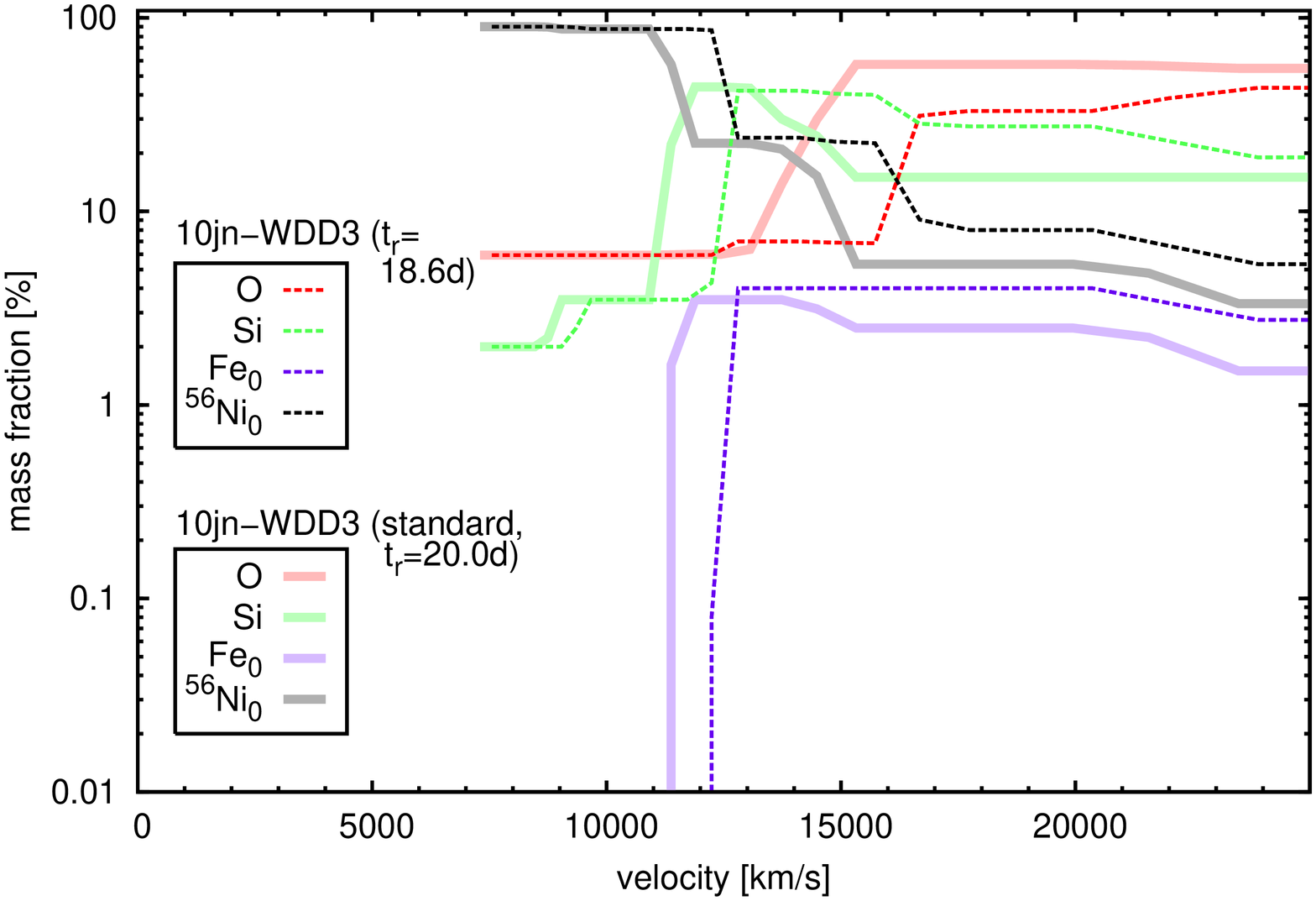}
  \caption{Distribution of the most abundant\,/\,relevant elements (in velocity
  space) in our $\tau_r= \textrm{18.6}$\,{}d models compared to the those in
  the standard ($\tau_r= \textrm{20}$\,{}d) models. \textit{Top panel:} 
  W7-based model; \textit{bottom panel:} WDD3-based model. The black and violet 
  graphs [mass fractions of \Nifs\ and stable Fe at $t=\textrm{0}$: 
  $X(^{56}\textrm{Ni}_0)$, $X(\textrm{Fe}_0)$] give the abundances of the
  dominant iron-group elements, which are unrealistically high in the outer 
  layers of the \mbox{$\tau_r=\textrm{18.6}$\,{}d} models (see text).}
  \label{fig:abundances-10jn-18-20-d}
\end{figure}

In order to calculate the models presented in the main text we have adopted a value of 20\,d for the $B$-band rise time (in the rest frame of the SN). Here we discuss why we have chosen this rise time, which is a bit longer than the one which is inferred from the observations assuming a $t^2$ behaviour for the rising part of the light curve ($\tau_{r,\textrm{fire}}= \textrm{18.6}$\,d; Sec. \ref{sec:observations-lcprop}).

During the process of modelling the SN, we have created various test models in order to determine which rise time allows for optimum models in terms of fit quality and physical consistency. This is motivated by the fact that the rise time of SNe Ia is still somewhat uncertain. Statistical studies in the late 1990s and early 2000s yielded somewhat different and controversial results for `average' SNe Ia [17.6\,d, \citet{groom98}; 19.5\,d, \citet{rie99}; 19.1\,/\,19.6\,d, \citet{con06}; 17.4\,d, \citet{strovink07}]. With current studies \citep[\eg][]{hayden10a,ganeshalingham11}, it has become clear that the rise time is correlated with the decline rate of the SN, \ie slowly declining objects like SN\,2010jn [\Dm\,$=$\,0.9] usually need a significantly longer time to reach their $B$-band maximum. Thus, SNe Ia are now known to have rise times between 13\myto{}23\,d \citep{hayden10a}, with an average of 17\myto{}18\,d  according to current estimates \citep{ganeshalingham11,gonzales-gaitan12}. Also, the shape of the early light-curve rise, which is usually assumed to be $\propto t^2$, probably shows variations from object to object due to differences in the \Nifs\ distribution, as recently emphasised by \citet{piro12b}.

In this Appendix, we discuss the differences between models assuming $\tau_{r,\textrm{fire}}= \textrm{18.6}$\,{}d and $\tau_r= \textrm{20}$\,{}d, which lead us to accept the latter value as the one more probable for SN\,2010jn. In the models with $\tau_r= \textrm{20}$\,d the ejecta have had more time to expand prior to the observations. The radius of a photosphere located at a certain velocity $v_\textrm{ph}$ is thus larger if we assume $\tau_{r}= \textrm{20}$\,{}d instead of $\tau_{r,\textrm{fire}}= \textrm{18.6}$\,{}d. This affects the spectrum at the photosphere: a smaller temperature $T_\textrm{ph}$ is sufficient to emit the same amount of radiation from a larger photosphere.

To have photospheres at the same radius in the  $\tau_r= \textrm{20}$\,d and $\tau_r= \textrm{18.6}$\,d models, keeping a similar colour and temperature, a higher $v_{ph}$  must be used for $\tau_r= \textrm{18.6}$\,d. For the earliest spectra ($-\textrm{12.9}$\,d, $-\textrm{10.5}$\,d), $v_{ph}$ in our standard (20\,d) model is already very high. An even higher velocity would mean that almost no mass is above the photosphere, but then line absorption would be reduced and it should be difficult to reproduce the low observed UV flux (\ie the flux in the model would be too high). The alternative would be to preserve the photospheric velocity from the standard calculation. In this case, however, the photospheric spectrum would be bluer (higher $T_\textrm{ph}$ required to emit the same luminosity from a small-radius photosphere, as mentioned above). Again, one would be left with too much UV flux.

The only way to circumvent this and fit the observed spectra with $\tau_r = \textrm{18.6}$\,d is to use higher mass fractions of the iron-group elements in the outer ejecta in order to block the UV flux. Our standard ($\tau_r = \textrm{20}$\,d) models have already shown a relatively strong out-mixing of iron-group material (Sections \ref{sec:tomography-10jn-w7-abundancestructure}, \ref{sec:tomography-10jn-wdd3-abundancestructure}) with respect to the original W7\,/\,WDD3 models \citep{iwa99}. Assuming $\tau_r = \textrm{18.6}$\,d, the iron-group fractions in the outer layers are even higher and less realistic. The most relevant abundances in the $\tau_r = \textrm{18.6}$\,d and $\tau_r = \textrm{20}$\,d models are compared in Fig. \ref{fig:abundances-10jn-18-20-d}. In the W7-based model with $\tau_r = \textrm{18.6}$\,d, an iron-group abundance (sum of all iron-group elements) of $\sim$20\% is needed even above 15000\,\kms. In the W7 nucleosynthesis calculations \citep{iwa99}, mass fractions of the order of 10\% are only reached further inwards (at $<$\,13010\,\kms), which constitutes an inconsistency. In general, it will be difficult for most explosion models to produce high abundances of iron-group material in the outermost layers: the densities there are small, leading to incomplete burning, and only `mixing' or multi-D effects in the abundance distribution can bring burned material outwards. Our WDD3-based model shows a smaller effect, but still the high iron-group abundance in the outer layers of the $\tau_r = \textrm{18.6}$\,d model ($\sim$\,14\% above 15000\,\kms) leads us to prefer a rise time of $\gtrsim \textrm{20}$\,d, for which the iron-group abundances in the outer layers of our models seem consistent with the WDD3 results if one assumes moderate additional out-mixing (\cf Sec. \ref{sec:tomography-10jn-wdd3-abundancestructure}).

Models for UV observations at even earlier epochs would react even more sensitively to the assumed rise time. If earlier UV observations were available for SN\,2010jn, a stricter lower limit on the rise time may have been determined through modelling. In order to exploit this possibility for future objects, we encourage UV observations of SNe\,Ia at epochs as early as possible.

\end{document}